\def\year{2021}\relax
\documentclass[letterpaper]{article} 
\usepackage{aaai21}  
\usepackage{times}  
\usepackage{helvet} 
\usepackage{courier}  
\usepackage[hyphens]{url}  
\usepackage{graphicx} 
\usepackage{csquotes}
\usepackage{rotating}
\usepackage{natbib}  
\usepackage{caption} 
\usepackage{array}
\usepackage{setspace} 
\usepackage{makecell}
\usepackage[flushleft]{threeparttable}
\usepackage{booktabs,caption}
\usepackage[toc,page]{appendix}
\usepackage{subcaption}
\usepackage{rotating}
\usepackage{multirow}

\title{State-level Racially Motivated Hate Crimes Contrast Public Opinion on the \#StopAsianHate and \#StopAAPIHate Movement}
\author{
    Hanjia Lyu, \textsuperscript{\rm 1}
    Yangxin Fan, \textsuperscript{\rm 1}
    Ziyu Xiong, \textsuperscript{\rm 1}
    Mayya Komisarchik, \textsuperscript{\rm 2}
    Jiebo Luo \textsuperscript{\rm 3,*}
    \\
}
\affiliations{
    \textsuperscript{\rm 1} University of Rochester, Goergen Institute for Data Science, Rochester, 14627, USA\\
    \textsuperscript{\rm 2} University of Rochester, Department of Political Science, Rochester, 14627, USA\\
    \textsuperscript{\rm 3} University of Rochester, Department of Computer Science, Rochester, 14627, USA\\

    \textsuperscript{\rm *} Corresponding author: Jiebo Luo (jluo@cs.rochester.edu)
}

\begin{document}

\maketitle

\begin{abstract}
\#StopAsianHate and \#StopAAPIHate are two of the most commonly used hashtags that represent the current movement to end hate crimes against the Asian American and Pacific Islander community. We conduct a social media study of public opinion on the \#StopAsianHate and \#StopAAPIHate movement based on 46,058 Twitter users across 30 states in the United States ranging from March 18 to April 11, 2021. The movement attracts more participation from women, younger adults, Asian and Black communities. 51.56\% of the Twitter users show direct support, 18.38\% are news about anti-Asian hate crimes, while 5.43\% show a negative attitude towards the movement. Public opinion varies across user characteristics. Furthermore, among the states with most racial bias motivated hate crimes, the negative attitude towards the \#StopAsianHate and \#StopAAPIHate movement is the weakest. To our best knowledge, this is the first large-scale social media-based study to understand public opinion on the \#StopAsianHate and \#StopAAPIHate movement. We hope our study can provide insights and promote research on anti-Asian hate crimes, and ultimately help address such a serious societal issue for the common benefits of all communities. 

\end{abstract}

A recent report found that in 16 of the America's largest cities, anti-Asian hate crimes have surged around 145\% in 2020\footnote{https://www.csusb.edu/sites/default/files/FACT\%20SHEET-\%20Anti-Asian\%20Hate\%202020\%20rev\%203.21.21.pdf}. On March 16, a series of mass shootings occurred at three spas or massage parlors in the metropolitan area of Atlanta, Georgia, United States\footnote{https://www.cnn.com/2021/03/17/us/robert-aaron-long-suspected-shooter/index.html}. Six out of eight victims who were killed were Asian women. Since then, many anti-Asian-violence rallies have been held across the world. The discussion and debate around \#StopAsianHate and \#StopAAPIHate which represent the social movement that aims to end hate crimes against Asian American and Pacific Islander communities, is becoming heated on social media platform like Twitter. 

Previous studies have investigated social movements like \#BlackLivesMatter~\cite{taylor1999collective,yang2016narrative,freelon2016beyond,gallagher2018divergent,ray2017ferguson}, \#HeForShe~\cite{dibrito2019reducing} and \#MeToo~\cite{jaffe2018collective, jagsi2018sexual,zarkov2018ambiguities,duong2020metoo}, however, few research are about the hate crimes against Asian American and Pacific Islander communities. To fill the gap in understanding the \#StopAsianHate and \#StopAAPIHate movement, we conduct a social media study of public opinion of 46,058 Twitter users across 30 states in the United States ranging from March 18 to April 11, 2021, and make four major contributions. To summarize, (1) using the publicly available self-disclosed information of Twitter users, our study analyzes the participation patterns in the movement; (2) based on the content and opinion of the tweets, we classify them into six major topics; (3) by conducting logistic regression, we find that public opinion on the \#StopAsianHate and \#StopAAPIHate movement varies across user characteristics; (4) we show that the rate of state-level racial bias motivated hate crimes is negatively associated with the proportion of tweets that show a negative attitude towards the movement. To our best knowledge, this is the first large-scale social media-based study to understand public opinion on the \#StopAsianHate and \#StopAAPIHate movement. 

\begin{table*}[htbp]
\caption{Descriptive statistics of the Twitter users in this study and the rate of state-level racial bias motivated hate crimes.}
     \label{tab:descriptive_vars}
    \begin{subtable}[t]{0.48\textwidth}
    \caption{Categorical}
       \label{tab:categorical}
        \centering
        \begin{tabular}{l l }
        \toprule
         & \textbf{n (\%)}\\
        \midrule
        Gender & \\
        Female & 25,646 (55.68)\\
        Male & 20,412 (44.32)\\
        Race/Ethnicity & \\
        {\tt White} & 24,746 (53.73) \\
        {\tt Black} & 8,950 (19.43) \\
        {\tt Hispanic} &  713 (1.55)  \\
        {\tt Asian} & 11,649 (25.29) \\
        Political affiliation  & \\
        Following Trump & 2,718 (5.90)\\
        Following Biden & 6,598 (14.32)\\
        Population density  & \\
        Urban & 33,526 (72.79)\\
        Suburban & 5,816 (12.63)\\
        Rural & 6,716 (14.58)\\
        Others  & \\
        Religious status & 3,046 (6.61)\\
        Family status & 3,171 (6.88)\\
        {\tt Verified} & 1,158 (2.51)\\
            \bottomrule
       \end{tabular}
       
    \end{subtable}
    \hfill
    \begin{subtable}[t]{0.48\textwidth}
    \caption{Continuous}
        \label{tab:continuous}
        \centering
        \begin{tabular}{l  l}
        \toprule
         & \textbf{Mean (SD)} \\
        \midrule
        Age & 34.76 (13.64) \\
        Income & 31,177 (8,894) \\
        \# of {\tt Friends} & 1,151 (4,718) \\
        \# of {\tt Followers} & 3,533 (149,239)\\
        \# of {\tt Listed memberships} & 32.45 (317.44) \\
        \# of {\tt Statuses} & 23,156 (45,301) \\
        \# of {\tt Favorites} & 34,211 (55,767) \\
        \makecell[l]{\# of racial bias motivated hate \\crimes per 10,000} & 0.13 (0.10)\\
        \bottomrule
        \end{tabular}
        
     \end{subtable}
     
\end{table*}

\section*{Participation patterns}
Using the Tweepy API and a list of related keywords and hashtags (see Materials and Methods), we have collected, inferred characteristics of 46,058 unique Twitter users across 30 states in U.S. ranging from March 18 to April 11, 2021, who have discussed the \#StopAsianHate and \#StopAAPIHate movement, and obtained the number of state-level racial bias motivated hate crimes from Federal Bureau of Investigation~\cite{fbi2019hate}, which are listed in Table~\ref{tab:descriptive_vars}. 55.68\% of the study population are women, which is greater than the proportion of women in the general Twitter population. According to the survey conducted by the Pew Research Center, women account for 50\% of the Twitter population~\cite{pew2019twitter}. This suggests that the \#StopAsianHate and \#StopAAPIHate movement attracts more participation from women, which might be related to collective identity~\cite{polletta2001collective} that women may join the movement based on shared membership of the victims of recent anti-Asian attacks where six out of eight people killed in Atlanta shootings were {\tt Asian} women, and so was the victim of the San Francisco attack\footnote{https://sanfrancisco.cbslocal.com/2021/03/17/elderly-asian-woman-beats-up-man-attacking-her-in-san-francisco/}. Similarly, {\tt Asian} users comprise 25.29\%, which is greater than the proportion in the general Twitter population (less than 8\%~\cite{pew2019twitter}). There are proportionally more {\tt Black} (19.43\%), fewer {\tt White} (53.73\%), and fewer {\tt Hispanic} Twitter users (1.55\%) involved in our study population than in the general Twitter population ({\tt Black}: 11\%, {\tt White}: 60\%, {\tt Hispanic}: 17\%)~\cite{pew2019twitter}. The Twitter users of our study population are relatively younger. In particular, 46.96\% are adults who are between 18 and 29 years old, 37.08\% are between 30 and 49, while in the general Twitter population, only 29\% for 18-29, and 44\% for 30-49~\cite{pew2019twitter}. This pattern is consistent with previous study that older adults are not only lagging behind in terms of physical access to the Internet but also in engaging in political activities in the online environment~\cite{xie2008older}.

\begin{figure*}[htbp]
     \centering

     \begin{subfigure}[b]{0.48\textwidth}
         \centering
         \includegraphics[width=\textwidth]{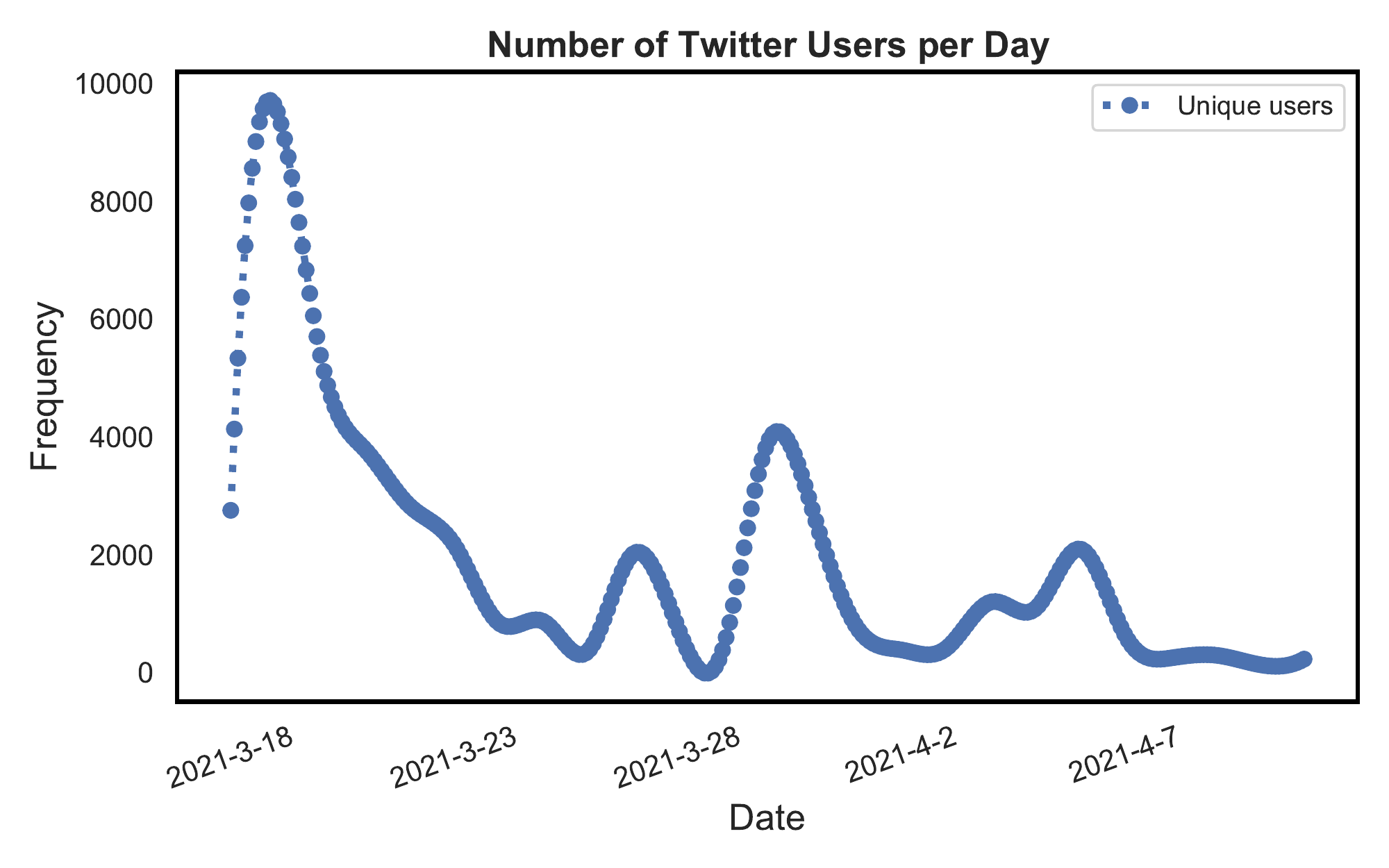}
         \caption{}
         \label{fig:users_per_day}
     \end{subfigure}
     \hfill
     \begin{subfigure}[b]{0.48\textwidth}
         \centering
         \includegraphics[width=\textwidth]{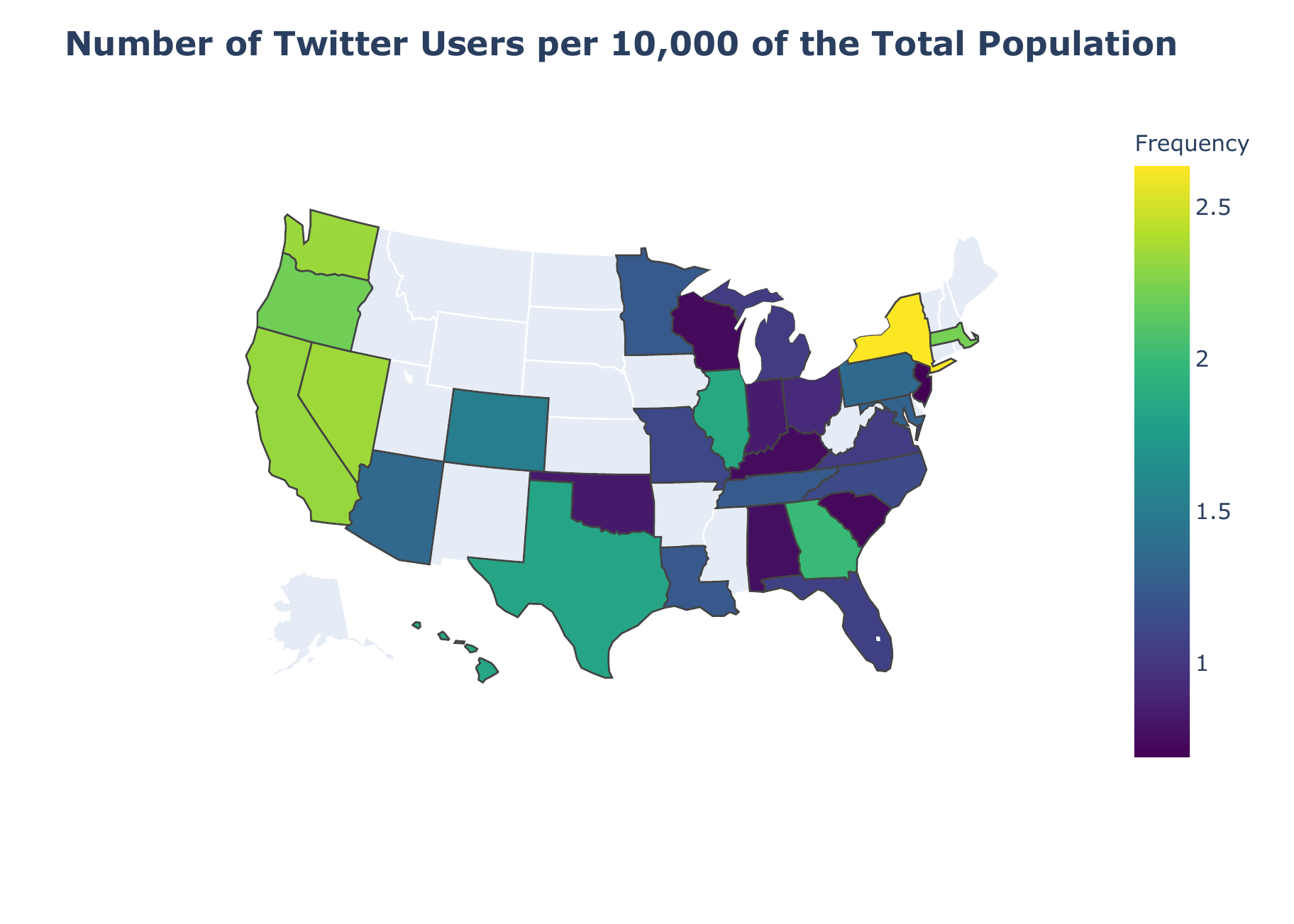}
          \vspace{-1cm}
         \caption{}
         \label{fig:geo_dist}
     \end{subfigure}
     
        \caption{(a) Number of Twitter users per day. (b) Number of Twitter users per 10,000 of the total population.}
        \label{fig:temporal_spatial}
\end{figure*}

As shown in Figure~\ref{fig:users_per_day}, the discussion about \#StopAsianHate and \#StopAAPIHate was most heated on March 19. After that, the number of unique Twitter users decreased gradually. On March 31, there was a surge again, where the most retweeted tweet is the news about an attack against a 65-year-old Asian American woman\footnote{https://www.nytimes.com/2021/03/30/nyregion/asian-attack-nyc.html}. Figure~\ref{fig:geo_dist} illustrates the state-level relative frequency of Twitter users who participate in the \#StopAsianHate and \#StopAAPIHate movement, which roughly correspond to the {\tt Asian} population percentage by state, where the West coast states, Illinois, Texas, Hawaii, and New York are the states with the highest {\tt Asian} population percentage. There are also more users in Georgia which is within expectation due to the Atlanta mass shootings.

\section*{Topic encoding}
To better understand public opinion on the \#StopAsianHate and \#StopAAPIHate movement, we read the tweets that have been retweeted for at least 20 times, and classify them into two levels of topics. There are 6 level 1 topics including Support, News, Denouncement, Double standard, Negative opinion, and Policy. In addition, there are 43 level 2 topics (see Supplemental Materials and Methods). Table~\ref{tab:topic_dist} shows the breakdown of level 1 topics. The breakdown of level 2 topics are shown in the Supplemental Materials and Methods.

\begin{table*}[htbp!]
\centering
\begin{threeparttable}
\caption{Level 1 topic distribution by user characteristics.}
    \label{tab:topic_dist}
    
    \begin{tabular}{l l l l l l l}
    \toprule
         & Support & News & Denouncement & Double standard & Negative opinion & Policy \\
     & (1) & (2) & (3) & (4) & (5) & (6) \\
         \midrule
        \textbf{Total} &  51.56\% & 18.38\% & 14.69\% & 8.37\% & 5.43\% & 1.56\%\\
        Male &  46.18\% & 21.51\% & 15.04\% & 9.83\% & 6.08\% & 1.37\% \\
        Female & 55.84\% & 15.89\% & 14.42\% & 7.21\% & 4.92\% & 1.72\%\\
        Age (18-29) & 54.24\% & 15.31\% & 12.96\% & 9.36\% & 6.90\% & 1.23\%\\
        Age (30-49) & 51.00\% & 19.74\% & 15.18\% & 7.84\% & 4.58\% & 1.66\%\\
        Age (50-64) & 46.44\% & 23.25\% & 18.05\% & 6.72\% & 3.30\% & 2.25\%\\
        Age (65+) & 40.90\% & 27.01\% & 20.45\% & 6.66\% & 2.54\% & 2.44\%\\
        {\tt Asian} & 58.96\% & 17.43\%& 12.56\% & 6.26\%& 3.92\%&0.87\%   \\
        {\tt White} &  50.74\% & 20.34\%& 16.56\%& 7.06\%& 3.21\%& 2.08\%\\
        {\tt Black} & 43.97\%& 14.17\%& 12.29\%& 14.86\%& 13.71\%& 1.01\%\\
        {\tt Hispanic} & 54.56\%& 18.65\% & 14.87\% &6.87\% & 3.23\% & 1.82\%\\
        Following Trump &37.05\% &28.37\%&16.89\%&9.16\%&7.14\%&1.40\% \\
        Following Biden & 55.18\% &20.25\%&16.91\%&3.64\%&1.23\%&2.79\%\\
    \bottomrule
    
    \end{tabular}
    \end{threeparttable}
\end{table*}

\subsubsection*{Support} 
The majority of the tweets are direct support which comprise 51.56\%. People post direct supportive statement to support the \#StopAsianHate and \#StopAAPIHate movement, specific groups and people. For instance, one of the most retweeted tweets with direct supportive statement is: ``\#StopAsianHate Pass it on.'' In addition, tweets that are attached with links to external resources like funding campaigns for victims, and documentary about Asian American history are prevalent in our study. The support from political leaders and celebrities are included in this category as well. Furthermore, there are tweets that talk about or encourage solidarity between {\tt Black} and {\tt Asian} communities. Besides that, Asian Americans post ``Proud to be Asian.'' to build solidarity within the {\tt Asian} community, which can be supported by previous study that boundary-setting rituals and institutions that separate them from whose in power can strengthen internal solidarity~\cite{taylor1999collective}. 

\subsubsection*{News}
This topic accounts for 18.38\%. The tweets of this category are the news about anti-Asian crime/talk. For some of them, the racial identity of the offender is not disclosed, while for some of them, the tweets explicitly state the racial identity of the offender.

\subsubsection*{Denouncement}
The third most tweets are denouncement. It is noteworthy that these tweets are not against \#StopAsianHate or \#StopAAPIHate. People express positive opinion on the \#StopAsianHate and \#StopAAPIHate movement by denouncing specific people/groups or systematic problems. For instance, one of the most retweeted tweets criticizes Donald Trump for calling COVID-19 as ``China virus''. Some tweets denounce the Republican party for their vote against the Violence Against Women Act. One of the Democratic political figure is criticized for spreading misinformation. Apart from the denouncement of people/groups, there is denouncement of racism (white supremacy), culture (Asian fetishization) and education (college admission).

\subsubsection*{Double standard}
8.37\% of the users discuss the double standard issues about media bias and different treatment. For instance, some tweets argue that {\tt Asian} and {\tt Black} are treated differently: ``Asian American `survivors' of assault are being allocated 50 million dollars???? $< url >$'' Some tweets claim that {\tt White} and {\tt Black} offenders are treated differently:``Why aren’t the two girls who murdered Asian immigrant Mohammad Anwar going to spend any time in jail?''

\subsubsection*{Negative opinion}
Representing 5.43\% of the Twitter users in our study population, this topic expresses negative opinion against the \#StopAsianHate and \#StopAAPIHate movement, including encouraging tension between {\tt Black} and {\tt Asian} communities, inciting anti-Asian sentiment, and justifying that hate is not the reason for the attacks.

\subsubsection*{Policy}
Accounting for 1.56\% users, the tweets of this topic explicitly demand for policy change or make references to specific laws. For instance, there are tweets that refer to gun control: ``In the aftermath of the anti-Asian attacks in Atlanta, just a reminder: Waiting period for an abortion in GA: 24 hours Waiting period for a gun in GA: none''.

\section*{Public opinion varies across user characteristics}
We conduct logistic regression to examine the predictive effect of user characteristics and the state-level rate of racial bias motivated hate crimes on the choice of level 1 and level 2 topics. The results of logistic regression of six level 1 topics are summarized in Table~\ref{tab:level1_logit_sum}. Each column represents a logistic regression model. The complete logistic regression outputs for level 1 and level 2 topics are further presented in the Supplemental Materials and Methods.

\begin{table*}[htbp!]
\small
\centering
\begin{threeparttable}
\caption{Logistic regression outputs for the opinion on the \#StopAsianHate and \#StopAAPIHate movement against demographics and other variables of interest.}
    \label{tab:level1_logit_sum}
    \centering
    
    \begin{tabular}{l l l l l l l}
    \toprule
    Independent variable     & Support & News & Denouncement & Double standard & Negative opinion & Policy \\
     & (1) & (2) & (3) & (4) & (5) & (6)\\
         \midrule
        Male & \makecell[l]{-$0.27^{***}$ \\ (0.02)} & \makecell[l]{$0.30^{***}$ \\ (0.03)} & \makecell[l]{0.01 \\ (0.03)} & \makecell[l]{$0.28^{***}$ \\ (0.04)} & \makecell[l]{$0.14^{**}$ \\ (0.04)}& \makecell[l]{-$0.23^{**}$\\(0.08)  }\\
        
        Age (years) & \makecell[l]{-$0.01^{***}$ \\ (0.00)} &  \makecell[l]{$0.01^{***}$ \\ (0.00)} & \makecell[l]{$0.01^{***}$ \\ (0.00)} & \makecell[l]{-0.00 \\ (0.00)} & \makecell[l]{-$0.01^{***}$ \\ (0.00)} & \makecell[l]{$0.01^{**}$\\(0.00)}\\
        
        {\tt White} & \makecell[l]{-$0.25^{***}$ \\ (0.02)} &  \makecell[l]{0.06 \\ (0.03)} & \makecell[l]{$0.27^{***}$ \\ (0.03)} & \makecell[l]{$0.17^{***}$ \\ (0.05)} & \makecell[l]{-0.06 \\ (0.06)} & \makecell[l]{$0.84^{***}$ \\ (0.11)} \\
        
        {\tt Black} & \makecell[l]{-$0.46^{***}$ \\ (0.03)} &  \makecell[l]{-$0.34^{***}$ \\ (0.04)} & \makecell[l]{0.02 \\ (0.04)} & \makecell[l]{$0.72^{***}$ \\ (0.05)} & \makecell[l]{$1.10^{***}$ \\ (0.06)} & \makecell[l]{$0.36^{*}$ \\ (0.15)}\\
        
        {\tt Hispanic} & \makecell[l]{-0.12 \\ (0.08)} &  \makecell[l]{0.01 \\ (0.10)} & \makecell[l]{$0.22^{*}$ \\ (0.11)} & \makecell[l]{0.02 \\ (0.16)} & \makecell[l]{-0.23 \\ (0.22)} & \makecell[l]{$0.84^{**}$ \\ (0.30)}\\
        Following Trump & \makecell[l]{-$0.59^{***}$ \\ (0.04)} &  \makecell[l]{$0.51^{***}$ \\ (0.05)} & \makecell[l]{0.09 \\ (0.05)} & \makecell[l]{$0.22^{**}$ \\ (0.07)} & \makecell[l]{$0.61^{***}$ \\ (0.08)} & \makecell[l]{-0.19 \\ (0.17)}\\
        
        Following Biden & \makecell[l]{$0.24^{***}$ \\ (0.03)} &  \makecell[l]{-0.01 \\ (0.04)} & \makecell[l]{0.07 \\ (0.04)} & \makecell[l]{-$0.81^{***}$ \\ (0.07)} & \makecell[l]{-$1.26^{***}$ \\ (0.12)} & \makecell[l]{$0.56^{***}$ \\ (0.09)}\\
        
        Hate crimes & \makecell[l]{0.16 \\ (0.10)} &  \makecell[l]{-0.02 \\ (0.13)} & \makecell[l]{0.23 \\ (0.14)} & \makecell[l]{-0.06 \\ (0.19)} & \makecell[l]{-$1.80^{***}$ \\ (0.27)}&\makecell[l]{0.22 \\ (0.38)}\\
        
        
        
        
        N &  \multicolumn{6}{c}{46,085}\\
    \bottomrule
    
    \end{tabular}
    \begin{tablenotes}
    \small
    \item Note. * $p<0.05$. ** $p<0.01$. *** $p<0.001$. Each model includes control variables (see Supplemental Materials and Methods). Table entries are coefficients (standard errors).
    \end{tablenotes}
    \end{threeparttable}
\end{table*}

\subsubsection*{Women are more likely to state direct support and demand for policy change} The majority tweets of men and women are about showing support. However, there are some nuanced differences. 55.84\% tweets of women are about direct support while this topic only accounts for 46.18\% of the tweets by men. By conducting logistic regression (summarized in Table~\ref{tab:level1_logit_sum}), we find that women retweet more direct supportive tweets ($B=-0.27, SE = 0.02, p < .001, OR = 0.76, 95\%CI = [0.73, 0.79]$), more tweets that explicitly demand for policy change ($B=-0.23, SE = 0.08, p < .01, OR = 0.79, 95\%CI = [0.68, 0.93]$), fewer news ($B=0.30, SE = 0.03, p < .001, OR = 1.35, 95\%CI = [1.28, 1.42]$), fewer negative tweets against \#StopAsianHate and \#StopAAPIHate ($B=0.14, SE = 0.04, p < .01, OR = 1.15, 95\%CI = [1.05, 1.25]$), and fewer discussions about double standard ($B=0.28, SE = 0.04, p < .001, OR = 1.32, 95\%CI = [1.23, 1.42]$) than males.

\subsubsection*{Men discuss in a more general way, while women talk about more specific issues}
Men are more likely to retweet general supportive statement ($B=0.16, SE = 0.03, p < .001, OR = 1.17, 95\%CI = [1.12, 1.25]$), tweets about political figures' ($B=0.20, SE = 0.06, p < .01, OR = 1.28, 95\%CI = [1.08, 1.38]$) or celebrity's support ($B=0.38, SE = 0.05, p < .001, OR = 1.46, 95\%CI = [1.32, 1.60]$), and tweets supporting ``All lives matter.'' ($B=0.99, SE = 0.15, p < .001, OR = 2.53, 95\%CI = [2.0 1, 3.63]$) (Table~\ref{tab:level2_logit_support}). However, women tend to show support by retweeting tweets about more specific issues like resources ($B=-0.54, SE = 0.04, p < .001, OR = 0.58, 95\%CI = [0.54, 0.63]$), stories about the victims ($B=-0.50, SE = 0.10, p < .001, OR = 0.61, 95\%CI = [0.50, 0.73]$). For denouncement (Table~\ref{tab:level2_logit_denouncement}), men retweet more general discussion about anti-Asian racism ($B=0.15, SE = 0.07, p < .05, OR = 1.16, 95\%CI = [1.01, 1.34]$), while women talk about western imperialism ($B=-0.53, SE = 0.18, p < .01, OR = 0.59, 95\%CI = [0.41, 0.84]$), Asian fetishization ($B=-0.89, SE = 0.27, p < .01, OR = 0.41, 95\%CI = [0.24, 0.69]$), and issues about {\tt White} family with {\tt Asian} members ($B=-0.54, SE = 0.24, p < .05, OR = 0.58, 95\%CI = [0.36, 0.94]$). Interestingly, we also find significant difference for discussion about college admission, where men pay more attention ($B=1.13, SE = 0.29, p < .001, OR = 3.10, 95\%CI = [1.75, 5.42]$). For news (Table~\ref{tab:level2_logit_news}), women retweet the tweets that share personal experience involving harm or racism ($B=-0.44, SE = 0.06, p < .001, OR = 0.64, 95\%CI = [0.58, 0.72]$), while men retweet the anti-Asian harm news that is written by media ($B=0.28, SE = 0.05, p < .001, OR = 1.32, 95\%CI = [1.20, 1.45]$).

\subsubsection*{Older adults retweet more news, denouncement and demand for policy change}
If the age is to increase by one year, the adult is 1.01 times more likely to retweet news ($B=0.01, SE = 0.00, p < .001, OR = 1.01, 95\%CI = [1.01, 1.01]$), 1.01 times more likely to retweet make denouncement ($B=0.01, SE = 0.00, p < .001, OR = 1.01, 95\%CI = [1.01, 1.01]$) and 1.01 times more likely to retweet tweets seeking policy change ($B=0.01, SE = 0.00, p < .01, OR = 1.01, 95\%CI = [1.00, 1.01]$). Take the ``18-29'' and ``65+'' age groups as an example, news-related tweets account for 27.01\% for ``65+'', however, they only account for 15.31\% for ''18-29''. 20.45\% tweets of ``65+'' make denouncement, while only 12.96\% tweets of ``18-29'' are related to denouncement. The younger adults are more likely to retweet direct support ($B=-0.01, SE = 0.00, p < .001, OR = 0.99, 95\%CI = [0.99, 0.99]$), and express negative opinion ($B=-0.01, SE = 0.00, p < .001, OR = 0.99, 95\%CI = [0.99, 0.99]$).

\subsubsection*{Older adults tend to denounce political figures and groups}
Of all the level 2 topics of ``Denouncement'' (Table~\ref{tab:level2_logit_denouncement}), age is found to have significant effect on the choice of topics that denounce specific political figures and groups. The older the adults are, the more likely, they denounce Donald Trump ($B=0.03, SE = 0.00, p < .001, OR = 1.03, 95\%CI = [1.02, 1.04]$), Republican ($B=0.02, SE = 0.00, p < .001, OR = 1.02, 95\%CI = [1.02, 1.03]$) and Democratic parties ($B=0.05, SE = 0.01, p < .001, OR = 1.05, 95\%CI = [1.04, 1.05]$).

\subsubsection*{Race/ethnicity shows significant effect on the choice of topics}
Using {\tt Asian} as the reference group, we find {\tt White} adults are less likely to state direct support tweets ($B=-0.25, SE = 0.02, p < .001, OR = 0.78, 95\%CI = [0.74, 0.81]$), more likely to retweet denouncement ($B=0.27, SE = 0.03, p < .001, OR = 1.31, 95\%CI = [1.22, 1.39]$), double standard related tweets ($B=0.17, SE = 0.05, p < .001, OR = 1.19, 95\%CI = [1.08, 1.30]$), and demand for policy change ($B=0.84, SE = 0.11, p < .001, OR = 2.32, 95\%CI = [1.86, 2.89]$). 

{\tt Black} adults are less likely to state direct support tweets ($B=-0.46, SE = 0.03, p < .001, OR = 0.63, 95\%CI = [0.59, 0.67]$), and retweet news ($B=-0.34, SE = 0.04, p < .001, OR = 0.71, 95\%CI = [0.66, 0.77]$). They are more likely to express negative opinion ($B=1.10, SE = 0.06, p < .001, OR = 3.00, 95\%CI = [2.66, 3.39]$), discuss double standard ($B=0.72, SE = 0.05, p < .001, OR = 2.05, 95\%CI = [1.86, 2.27]$), and demand for policy change ($B=0.36, SE = 0.15, p < .05, OR = 1.43, 95\%CI = [1.07, 1.92]$).

{\tt Hispanic} adults are only found to be more likely to retweet denouncement ($B=0.22, SE = 0.11, p < .05, OR = 1.25, 95\%CI = [1.00, 1.54]$), and demand for policy change ($B=0.84, SE = 0.30, p < .01, OR = 2.32, 95\%CI = [1.30, 4.18]$).

\subsubsection*{Different race/ethnicity groups tend to defend themselves when discussing \#StopAsianHate and \#StopAAPIHate}  
{\tt Black} community is more likely to discuss solidarity between them and {\tt Asian} community ($B=0.89, SE = 0.10, p < .001, OR = 2.44, 95\%CI = [1.99, 2.94]$) (Table~\ref{tab:level2_logit_support}). When it comes to double standard (Table~\ref{tab:level2_logit_double}), they focus on the topics that {\tt Black} adults and {\tt Asian} adults are treated differently ($B=1.24, SE = 0.29, p < .001, OR = 3.46, 95\%CI = [1.95, 6.11]$). In addition, they tend to retweet tweets that criticize white supremacy ($B=0.63, SE = 0.10, p < .001, OR = 1.88, 95\%CI = [1.54, 2.27]$) (Table~\ref{tab:level2_logit_denouncement}). On the contrary, {\tt White} adults are more likely to retweet tweets that attribute the anti-Asian crimes to {\tt Black} community instead of white supremacy ($B=0.69, SE = 0.31, p < .05, OR = 1.99, 95\%CI = [1.11, 3.63]$).

\subsubsection*{Political divides extend to the choice of topics}
Following Trump and following Biden are both found to have significant effect on direct support, negative opinion, and double standard. Compared to people who do not follow Biden, those who follow Biden are more likely to show direct support ($B=0.24, SE = 0.03, p < .001, OR = 1.27, 95\%CI = [1.20, 1.34]$), but less likely to express negative opinion ($B=-1.26, SE = 0.12, p < .001, OR = 0.28, 95\%CI = [0.23, 0.36]$), and talk about double standard ($B=-0.81, SE = 0.07, p < .001, OR = 0.44, 95\%CI = [0.39, 0.51]$). By contrast, compared to people who do not follow Trump, those who follow Trump are less likely to show direct support ($B=-0.59, SE = 0.04, p < .001, OR = 0.55, 95\%CI = [0.51, 0.60]$), but more likely to express  negative opinion ($B=0.61, SE = 0.08, p < .001, OR = 1.84, 95\%CI = [1.56, 2.16]$), and talk about double standard ($B=0.22, SE = 0.07, p < .01, OR = 1.25, 95\%CI = [1.08, 1.43]$).

\subsubsection*{Disagreements lie in five sub-topics between Biden's and Trump's followers}
For Trump followers, they are more likely to retweet the opinion that {\tt Black} is to be blamed for the anti-Asian crimes ($B=1.13, SE = 0.24, p < .001, OR = 3.10, 95\%CI = [1.93, 4.95]$), denounce the anti-Asian racism within college admission ($B=0.76, SE = 0.33, p < .05, OR = 2.14, 95\%CI = [1.11, 4.10]$) (Table~\ref{tab:level2_logit_denouncement}), support ``All lives matter'' ($B=0.23, SE = 0.03, p < .001, OR = 1.26, 95\%CI = [1.20, 1.34]$) (Table~\ref{tab:level2_logit_support}), retweet news about {\tt Black} targeting {\tt Asian} ($B=0.94, SE = 0.11, p < .001, OR = 2.56, 95\%CI = [2.10, 3.16]$), but less likely to retweet news about personal experience ($B=-0.94, SE = 0.13, p < .001, OR = 0.39, 95\%CI = [0.30, 0.50]$) (Table~\ref{tab:level2_logit_news}), compared to those who do not follow Trump. Following Biden have opposite effect on these five topics against following Trump.

\section*{Hate crimes and public opinion on \#StopAsianHate and \#StopAAPIHate}
\subsubsection*{Negative opinion is the weakest in the states with most hate crimes}
\begin{figure}
    \centering
    \includegraphics[width = \linewidth]{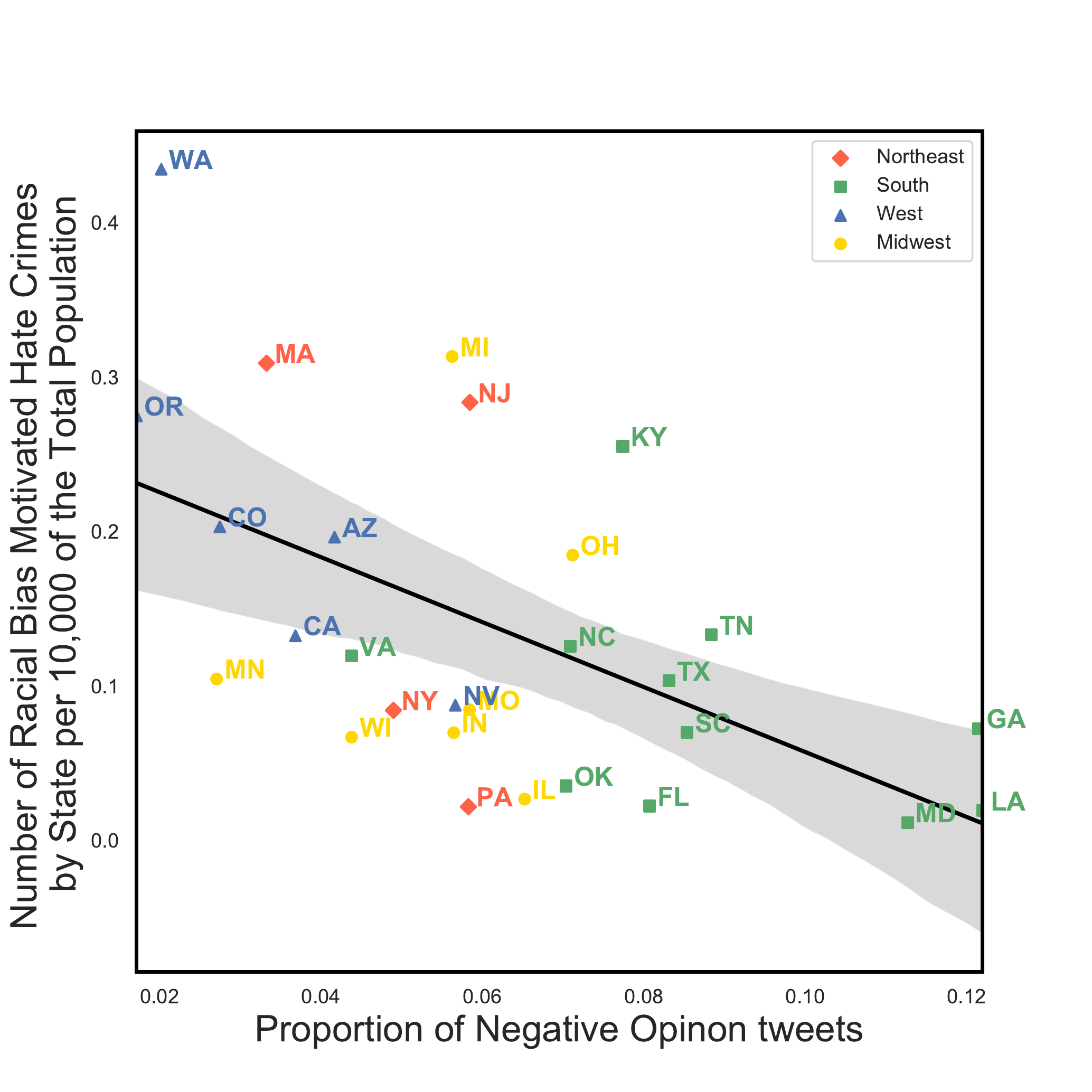}
    \caption{Number of racial bias motivated hate crimes by state per 10,000 of the total population versus proportion of negative opinion tweets.}
    \label{hate_vs_negative}
\end{figure}

The number of racial bias motivated hate crimes per 10,000 of the total population by state is negatively associated with the likelihood of expressing negative opinion ($B=-1.80, SE = 0.27, p < .001, OR = 0.16, 95\%CI = [0.10, 0.28]$). We further explore the relationship between hate crimes and topics of the tweets in Figure~\ref{hate_vs_negative}, where we color the states according to their geographic region. The states of the South are in the lower right corner. The clusters of the South, West, Midwest Northeast in Figure~\ref{hate_vs_negative} suggest a common opinion on anti-Asian hate crimes among spatially proximal populations.

\section*{Discussions}
The public opinion on the \#StopAsianHate and \#StopAAPIHate movement of 46,058 Twitter users across 30 states in the U.S. ranging from March 18 to April 11, 2021 are collected and analyzed. Among them, 51.56\% show direct support, 18.38\% retweet news about anti-Asian crimes, 14.69\% denounce anti-Asian racism, 8.37\% discuss issues about double standard, 1.56\% explicitly demand for policy change, while 5.43\% express negative opinion against the \#StopAsianHate and \#StopAAPIHate movement (Table~\ref{tab:topic_dist}). 

The \#StopAsianHate and \#StopAAPIHate movement is more likely to attract participation from women and younger adults. A similar pattern is observed in the \#BlackLivesMatter movement~\cite{olteanu2015characterizing}. The percentages of {\tt Asian} and {\tt Black} communities in our study population are greater than those in the general Twitter population. 

We observe that public opinion varies across user characteristics including gender, age, race/ethnicity as well as political affiliation. Women are more likely to state direct support and demand for policy change than men. They focus on specific issues while men tend to discuss in a more general way. Older adults retweet more news, denouncement and demand for policy change. Of all the level 2 topics of ``Denouncement'', older adults tend to denounce political figures and groups. Different race/ethnicity groups respond to topics about \#StopAsianHate and \#StopAAPIHate differently, and tend to defend themselves in the discussions. The {\tt Black} and {\tt White} communities point to each other as the ones to be blamed for the anti-Asian hate crimes. The creation of a strong movement identity leads to a backlash~\cite{polletta2001collective}. In our study, it is observed that when the {\tt Black} community criticizes white supremacy for the Atlanta attack, the {\tt White} community resists by retweeting tweets that attribute the anti-Asian crimes to the {\tt Black} community. For instance, one of the most retweeted references is the report of \textit{Criminal Victimization, 2018} by Bureau of Justice Statistics, where the data shows that {\tt Black}s are the major offenders of the violent incidents against {\tt Asian} community~\cite{morgan2019criminal}. Previous study suggests that {\tt White}s perceive their Whiteness as a negative attribute that now puts them at a perceptual disadvantage in society, and think that their Whiteness should not to be used as a marker of liability for continuing racial hate against minorities~\cite{jackson2002perceptions}. With respect to political affiliation, we find that political divides extend to the choice of topics in the \#StopAsianHate and \#StopAAPIHate movement, with Biden followers more likely to show direct support but less likely to express negative opinion or discuss double standard.

Furthermore, the rate of racial bias motivated hate crimes has a significant effect on negative opinion. In the places with the highest racial bias motivated hate crime rate, the negative opinion is the weakest. It is suggested that when individuals are targeted because of their race or ethnicity, they are likely to experience a host of negative emotions that are qualitatively distinct from those experienced following non-biased criminal victimization~\cite{craig2003after}. People who have not encountered hate crimes might express negative opinions because they do not have the related-experience, along with the negative emotions, thus making it harder for them to understand the seriousness of hate crimes. We further find that spatially proximal populations may share a common attitude towards anti-Asian crimes. To increase awareness and have a better understanding of racial bias motivated hate crimes, hate crime education programs can be deployed~\cite{anderson2002preventing}.

To our best knowledge, this is the first large-scale social media-based study to understand public opinion on the \#StopAsianHate and \#StopAAPIHate movement. We hope our work can provide insights and promote research on anti-Asian hate crimes, and ultimately help address such a serious societal issue for the common benefits of all communities. Future work could investigate the relationship between culture and public opinion.

\section*{Materials and Methods}
Full methods are in the Supplemental Materials and Methods.
\subsection*{Twitter Data}
We use the Tweepy API\footnote{(https://www.tweepy.org/)} to collect the related tweets which are publicly available. The search keywords and hashtags are related to the \#StopAsianHate and \#StopAAPIHate movement or {\tt Asian} community, including ``\#StopAsianHate'', ``\#StopAsianHateCrimes'', ``\#StopAAPIHate'', ``\#stopaapihate'', ``\#stopasianhate'', ``asian'', ``aapi'', ``\#asian'', ``\#aapi'', ``\#AAPI'', ``\#AsianLivesMatter'', ``\#AsiansAreHuman'', ``\#AntiAsianHate'', ``\#AntiAsianRacism'', ``\#JewsForAsian'', ``\#SafetyinSolidarity''.\footnote{(The non-hastag keywords are case-insensitive in the Tweepy query.)} Using a set of methods that are also applied in previous social media studies~\cite{lyu2020sense,lyu2020social,xiong2021gen,zhang2020influence}, we infer user characteristics including gender, age, race/ethnicity, income, religious status, family status, population density as well as political affiliation. Only the states that have at least 300 unique Twitter users are retained. 1,162 unique tweets from March 18 to April 11, 2021 that are retweeted for at least 20 times by 46,058 unique Twitter users are included in the study. 
\subsection*{Hate Crime Data}
We use the \textit{Hate Crime Statistics, 2019} from Federal Bureau of Investigation~\cite{fbi2019hate}, which comprises the hate crime incidents, per bias motivation and quarter, by State, Federal, and Agency of 2019. By the time this study is conducted, the data of 2020 and 2021 are not available yet.

\bibliography{main}

\clearpage

\appendix
\renewcommand\thefigure{\thesection.\arabic{figure}}    
\setcounter{figure}{0} \renewcommand{\thefigure}{A. \arabic{figure}}
\setcounter{table}{0} \renewcommand{\thetable}{A. \arabic{table}}
\setcounter{figure}{0}
\section{Supplemental Materials and Methods}

\subsection{Feature Inference}
\subsubsection{Gender and Age} Following the methods of \citet{lyu2020sense}, we use Face++ API\footnote{(https://www.faceplusplus.com/)} to infer the gender and age information of the users using their profile images. The invalid image URLs and images with multiple or zero faces are excluded. The gender and age information of the remaining users (i.e., there is only one intelligible face in the profile image) is inferred. Since our study focuses on the opinion of U.S. adults, the users who are younger than 18 are removed.

\subsubsection{Race/Ethnicity}
To estimate Twitter users’ race/ethnicity, we use both the text and image content. 

\noindent\textbf{Image-based Inference.} We use profile images from Twitter users and apply DeepFace API to analyze facial attributes~\cite{serengil2020lightface}. \\
\noindent\textbf{Text-based Inference.} We include user description and user name in this part of inference, and apply Ethnicolr API and CLD3 API for text analysis. 
\begin{itemize}
    \item \textbf{Description}: For each user description, we look for self-identified keywords, such as ``Chinese'',``Korean'' and ``Japanese'' for {\tt Asian}; ``African'', ``Algerian'' and ``Nigerian'' for {\tt Black}; ``Latino'' and ``Hispanic'' for {\tt Hispanic}.
    \item \textbf{User name}: We apply Ethnicolr API on user's last name. To extract the last name, we remove emoji icons, hyphens, unrelated contents and special characters; next, we split the remaining strings and keep the last part as the last name; lastly, we apply “census\_ln” from Ethnicolr to infer the race/ethnicity which contains {\tt White}, {\tt Black or African American}, {\tt Asian/Pacific Islander}, {\tt American Indian/Alaskan Native}, and {\tt Hispanic}.
    \item \textbf{Language detection on name}: We apply CLD3 API to user name, which can predict the language of the user name. 
\end{itemize}
There are four methods used in race/ethnicity inference (one for image-based inference, three for text-based inference), In order to give a final prediction of race/ethnicity, we decide to assign priorities for these four methods. Text-based description method has the highest priority, because we think the self-identified information is the most accurate factor among all these methods. Image-based method is the second. Ethnicolr API and CLD3 language detection are placed the third and fourth. In the end, we keep the users of {\tt White}, {\tt Black}, {\tt Hispanic} and {\tt Asian} in the study population and remove the race/ethnicity groups with few respondents.

\subsubsection{User-level Features}
Seven user-level features are crawled by the Tweepy API as well which include the number of {\tt Followers}, {\tt Friends}, {\tt Listed memberships}, {\tt Favourites}, {\tt Statuses}, the number of months since the user account was created, and the {\tt Verified} status. Moreover, we normalize the number of {\tt Followers}, {\tt Friends}, {\tt Listed memberships}, {\tt Favourites}, and {\tt Statuses} by the number of months since the user account was created.

\subsubsection{Geo-locations}
For Twitter, we choose to resolve the geo-locations using users' profiles. Similar to \citet{lyu2020sense}, the locations with noise are excluded, and the rest are classified into urban, suburban, or rural.

\subsubsection{Income}
The median household income of the census data at the place level is matched to represent the income of the Twitter user.

\subsubsection{Religious Status} We assign each user a Boolean value for whether he/she is religious based on the description in the profile~\cite{zhang2020influence}. 

\subsubsection{Family Status} By applying regular expression search, we identify users who show evidence that they are either fathers or mothers~\cite{zhang2020influence}. 

\subsubsection{Political Affiliations}
The political attribute is labelled based on whether this Twitter user followed the Twitter accounts of the top political leaders. The incumbent president (Joe Biden) and the former president (Donald Trump) are included in the analysis\footnote{(Due to the limitation of Twitter API, only about half of Donald Trump's follower IDs were crawled.)}.

    
    



\subsection{Topic Encoding}
To figure out what topics are discussed in the collected tweets, we first apply the Latent dirichlet allocation (LDA) topic modeling~\cite{blei2003latent} to acquire a general view of topic distribution. Next, we read and label the tweets that have been retweeted for at least 20 times. 

\begin{figure}[htbp!]
    \centering
    \includegraphics[width = \linewidth]{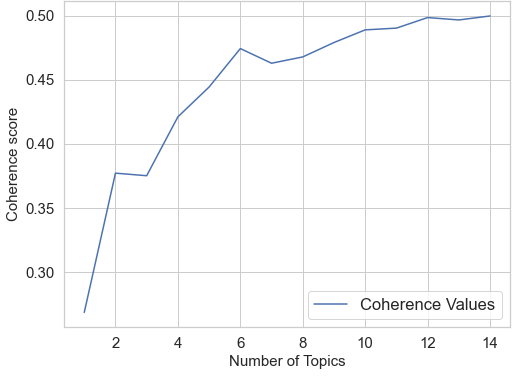}
    \caption{LDA coherence scores versus number of topics.}
    \label{fig:lda_coherence}
\end{figure}

\noindent\textbf{LDA Topic Modeling.} 
Based on the trend of coherence scores (Figure~\ref{fig:lda_coherence}), we choose 12 topics (Table~\ref{tab:lda_topics_summary}), which include Love, Hate crime, Racial problems, News about anti-Asian attacks, Violent crime, Community support, Individual support, Call for change, History, Politics, Culture, and Irrelevant. The results of the LDA guide us to further explore topics through manual labeling. 

\begin{table*}[htbp!]
\caption{Top 10 keywords of each topic.}
    \label{tab:lda_topics_summary}
    \centering
    \begin{tabular}{c c c}
    \toprule
        Topic & Keywords \\
         \midrule
        Love & love, friend, family, good, time, today, share, watch, hope, post \\
        Hate crime & hate, Asian, crime, anti, stop, racism, violence, year, rise, report\\
        Racial problems & people, white, black, Asian, race, supremacy, racist, problem, medium\\
        News about anti-Asian attacks & American, Asian, attack, Atlanta, shooting, victim, week, news, Georgia, Biden\\
        Violent crime & women, man, kill, day, bad, murder, target, shoot, sex, police\\
        Community support & community, violence, support, stand, racism, AAPI, solidarity, act, fight, action\\
        Individual support & Asian, people, happen, feel, life, speak, tweet, live, word, care\\
        Call for change & Asian, work, experience, face, time, hard, speak, learn, kid, change\\
        History & Asian, read, history, story, long, show, part, important, thread, write\\
        Politics & Asian, country, call, china, American, Chinese, America, Trump, virus, Covid\\
        Culture & Asian, make, culture, thing, talk, lot, point, comment, good, Japanese\\
        Irrelevant & Asian, racist, fuck, guy, shit, give, girl, fucking, back, big\\
    \bottomrule
    \end{tabular}
    
\end{table*}

\noindent\textbf{Labeling.}
Two researchers read the 1,335 tweets that have been retweeted for at least 20 times, and construct the topic classification as shown in Table~\ref{tab:topic_definition}. There are 6 level 1 topics, 43 level 2 topics, and 1 category with 173 irrelevant tweets (excluded from the study population). Next, they read the tweets and make judgement of the topic of each tweet independently. The label of the tweet is assigned based on mutual agreement.

\begin{table*}[htbp!]
\small
\centering
\begin{threeparttable}
\caption{Level 1 and level 2 topics.}
    \label{tab:topic_definition}
    \centering
    
\begin{tabular}{lll}
\toprule
Level 1                           & Level 2                                                                               & Code \\
\midrule
\multirow{9}{*}{Support}          & General supportive statement                                                          & S.1  \\
                                  & Resources                                                                             & S.2  \\
                                  & Stories of victims                                                                 & S.3  \\
                                  & Political figures' support                                                            & S.4  \\
                                  & Celebrity's support                                                                   & S.5  \\
                                  & Historical references                                                                 & S.6  \\
                                  & Solidarity between {\tt Black} and {\tt Asian}                                                    & S.7  \\
                                  & Proud to be {\tt Asian}                                                                     & S.8  \\
                                  & All lives matter                                                                      & S.9  \\
                                  \midrule
\multirow{6}{*}{News}             & News about anti-Asian hate crime / talk (no racial identity of the offender) & N.1  \\
                                  & News about {\tt Black} targeting {\tt Asian}                                                      & N.2  \\
                                  & News about {\tt White} targeting {\tt Asian}                                                      & N.3  \\
                                  & Personal experience                                                                   & N.4  \\
                                  & News about the support for \#StopAsianHate and \#StopAAPIHate                         & N.5  \\
                                  & News or reports that explicitly connect the pandemic to anti-Asian attacks             & N.6  \\
                                  \midrule
\multirow{17}{*}{Denouncement}    & Against Donald Trump                                                                  & D.1  \\
                                  & Against the Republican party                                                          & D.2  \\
                                  & Against the officer who justified for the offender of Atlanta shootings               & D.3  \\
                                  & Against the bystanders                                                                & D.4  \\
                                  & Against the Democratic party                                                          & D.5  \\
                                  & Politics and discrimination                                                           & D.6  \\
                                  & Against Kamala Harris                                                                 & D.7  \\
                                  & College admission                                                                     & D.8  \\
                                  & Asian fetishization                                                                   & D.9  \\
                                  & White supremacy                                                                       & D.10 \\
                                  & Western imperialism                                                                   & D.11 \\
                                  & {\tt Black} is the major offender                                                           & D.12 \\
                                  & Victims and offenders are treated differently by police                               & D.13 \\
                                  & {\tt White} family with {\tt Asian} members                                                       & D.14 \\
                                  & Against the attacker                                                                  & D.15 \\
                                  & General discussion about anti-Asian racism                                            & D.16 \\
                                  & Motivation of the Atlanta shootings is hate                                           & D.17 \\
                                  \midrule
\multirow{3}{*}{Double standard}  & Media bias                                                                            & DB.1 \\
                                  & {\tt Asian} and {\tt Black} are treated differently                                               & DB.2 \\
                                  & {\tt White} and {\tt Black} are treated differently                                               & DB.3 \\
                                  \midrule
\multirow{6}{*}{Negative opinion} & Tension between {\tt Black} and {\tt Asian}                                                       & NO.1 \\
                                  & Bystanders are not to be blamed                                                       & NO.2 \\
                                  & Motivation of the Atlanta shootings is not hate                                       & NO.3 \\
                                  & The offender is a crisis actor                                                                          & NO.4 \\
                                  & The fight between the {\tt Asian} man and the {\tt Black} man is not about hate                   & NO.5 \\
                                  & Incite anti-Asian sentiment                                                           & NO.6 \\
                                  \midrule
\multirow{2}{*}{Policy}           & Demand for policy change in general                                                   & P.1  \\
                                  & Gun control                                                                           & P.2 
\\
\bottomrule
\end{tabular}
    \end{threeparttable}
\end{table*}

\subsection{Level 2 topic distributions by user characteristics}
In this section, we show level 2 topic distributions by user characteristics including ``Support'' (Table~\ref{tab:level2_topic_support}), ``News'' (Table~\ref{tab:level2_topic_news}), ``Denouncement'' (Table~\ref{tab:level2_topic_denouncement}), ``Double standard'' (Table~\ref{tab:level2_topic_double}), ``Negative opinion'' (Table~\ref{tab:level2_topic_negative}), and ``Policy'' (Table~\ref{tab:level2_topic_policy}). 

\begin{table*}[htbp!]
\centering
\begin{threeparttable}
\caption{Level 2 topic distribution of ``Support'' by user characteristics.}
    \label{tab:level2_topic_support}
    \centering
\begin{tabular}{llllllllll}
\toprule
                & S.1     & S.2     & S.3    & S.4     & S.5     & S.6    & S.7    & S.8     & S.9    \\
                \midrule
\textbf{Total}           & 52.04\% & 16.00\% & 2.22\% & 5.20\%  & 8.89\%  & 6.73\% & 2.82\% & 5.16\%  & 0.93\% \\
Male            & 56.01\% & 11.03\% & 1.56\% & 6.38\%  & 10.61\% & 6.16\% & 2.59\% & 4.07\%  & 1.59\% \\
Female          & 49.43\% & 19.27\% & 2.65\% & 4.43\%  & 7.76\%  & 7.11\% & 2.97\% & 5.87\%  & 0.50\% \\
Age (18-29)     & 47.24\% & 19.02\% & 1.88\% & 2.58\%  & 11.04\% & 8.65\% & 3.44\% & 5.51\%  & 0.63\% \\
Age (30-49)     & 55.49\% & 15.03\% & 2.59\% & 5.56\%  & 7.88\%  & 5.81\% & 2.70\% & 3.83\%  & 1.11\% \\
Age (50-64)     & 60.18\% & 8.80\%  & 2.36\% & 12.44\% & 4.48\%  & 2.48\% & 1.24\% & 6.64\%  & 1.40\% \\
Age (65+)       & 59.58\% & 4.73\%  & 2.49\% & 17.16\% & 2.24\%  & 1.99\% & 0.12\% & 9.70\%  & 1.99\% \\
{\tt Asian   }        & 41.58\% & 15.75\% & 1.94\% & 3.16\%  & 10.32\% & 6.22\% & 2.90\% & 17.84\% & 0.29\% \\
{\tt White}           & 56.07\% & 18.18\% & 2.70\% & 6.78\%  & 6.51\%  & 6.56\% & 1.74\% & 0.00\%  & 1.47\% \\
{\tt Black }          & 57.26\% & 9.35\%  & 1.27\% & 3.84\%  & 13.70\% & 8.03\% & 6.28\% & 0.00\%  & 0.28\% \\
{\tt Hispanic}        & 54.24\% & 16.97\% & 1.03\% & 4.37\%  & 11.83\% & 8.48\% & 1.29\% & 0.00\%  & 1.80\% \\
Following Trump & 55.11\% & 9.83\%  & 2.18\% & 8.64\%  & 6.75\%  & 5.26\% & 1.89\% & 4.87\%  & 5.46\% \\
Following Biden & 59.96\% & 8.90\%  & 2.94\% & 13.13\% & 4.06\%  & 4.01\% & 1.24\% & 5.49\%  & 0.27\%\\
\bottomrule
\end{tabular}
    \end{threeparttable}
\end{table*}

\begin{table*}[htbp!]
\centering
\begin{threeparttable}
\caption{Level 2 topic distribution of ``News'' by user characteristics.}
    \label{tab:level2_topic_news}
    \centering

\begin{tabular}{lllllll}
\toprule
                & N.1     & N.2     & N.3    & N.4     & N.5     & N.6    \\
                \midrule
\textbf{Total}        & 62.31\% & 8.23\%  & 1.29\% & 22.03\% & 5.55\%  & 0.59\% \\
Male            & 65.83\% & 9.73\%  & 0.98\% & 16.99\% & 5.95\%  & 0.52\% \\
Female          & 58.51\% & 6.62\%  & 1.62\% & 27.45\% & 5.13\%  & 0.66\% \\
Age (18-29)     & 57.90\% & 5.19\%  & 2.08\% & 27.09\% & 7.31\%  & 0.42\% \\
Age (30-49)     & 65.45\% & 7.74\%  & 0.92\% & 19.57\% & 5.55\%  & 0.77\% \\
Age (50-64)     & 65.89\% & 12.38\% & 0.64\% & 18.13\% & 2.32\%  & 0.64\% \\
Age (65+)       & 61.39\% & 20.53\% & 0.19\% & 15.25\% & 2.26\%  & 0.38\% \\
{\tt Asian  }         & 60.91\% & 4.48\%  & 0.89\% & 28.26\% & 4.83\%  & 0.64\% \\
{\tt White }          & 63.63\% & 10.39\% & 1.37\% & 20.30\% & 3.68\%  & 0.64\% \\
{\tt Black}           & 59.86\% & 5.13\%  & 1.26\% & 19.24\% & 14.27\% & 0.24\% \\
{\tt Hispanic}        & 57.14\% & 13.53\% & 4.51\% & 18.80\% & 4.51\%  & 1.50\% \\
Following Trump & 64.46\% & 20.62\% & 1.30\% & 9.86\%  & 3.11\%  & 0.65\% \\
Following Biden & 69.54\% & 2.62\%  & 0.52\% & 23.58\% & 2.84\%  & 0.90\%\\
\bottomrule
\end{tabular}

    \end{threeparttable}
\end{table*}

\begin{sidewaystable*}[htbp!]
\centering
\scriptsize
\begin{threeparttable}
\caption{Level 2 topic distribution of ``Denouncement'' by user characteristics.}
    \label{tab:level2_topic_denouncement}
    \centering

\begin{tabular}{llllllllllllllllll}
\toprule
                & D.1     & D.2     & D.3    & D.4     & D.5     & D.6     & D.7    & D.8    & D.9    & D.10    & D.11   & D.12   & D.13    & D.14   & D.15   & D.16    & D.17    \\
                \midrule
\textbf{Total}           & 6.93\%  & 5.61\%  & 1.18\% & 15.66\% & 3.25\%  & 12.94\% & 0.03\% & 1.03\% & 1.26\% & 17.26\% & 2.23\% & 1.68\% & 5.20\%  & 1.24\% & 1.36\% & 14.89\% & 8.23\%  \\
Male            & 7.82\%  & 6.03\%  & 1.21\% & 15.09\% & 4.43\%  & 16.00\% & 0.00\% & 1.73\% & 0.62\% & 16.26\% & 1.47\% & 2.61\% & 3.03\%  & 0.85\% & 1.73\% & 15.09\% & 6.06\%  \\
Female          & 6.19\%  & 5.27\%  & 1.16\% & 16.14\% & 2.27\%  & 10.41\% & 0.05\% & 0.46\% & 1.78\% & 18.09\% & 2.87\% & 0.92\% & 7.00\%  & 1.57\% & 1.05\% & 14.73\% & 10.03\% \\
Age (18-29)     & 3.53\%  & 3.28\%  & 1.61\% & 14.20\% & 0.89\%  & 10.74\% & 0.00\% & 0.68\% & 1.93\% & 21.23\% & 3.32\% & 0.79\% & 8.96\%  & 1.32\% & 1.43\% & 17.63\% & 8.46\%  \\
Age (30-49)     & 6.83\%  & 5.90\%  & 0.81\% & 16.40\% & 2.82\%  & 15.08\% & 0.00\% & 1.04\% & 1.08\% & 15.93\% & 1.97\% & 1.81\% & 3.40\%  & 1.62\% & 1.74\% & 15.08\% & 8.49\%  \\
Age (50-64)     & 13.58\% & 9.67\%  & 1.13\% & 17.70\% & 8.23\%  & 13.99\% & 0.21\% & 1.65\% & 0.21\% & 11.01\% & 0.62\% & 3.29\% & 1.23\%  & 0.41\% & 0.62\% & 9.77\%  & 6.69\%  \\
Age (65+)       & 15.17\% & 10.20\% & 0.75\% & 16.17\% & 10.45\% & 11.94\% & 0.00\% & 1.99\% & 0.25\% & 13.18\% & 0.25\% & 3.23\% & 0.25\%  & 0.25\% & 0.25\% & 6.97\%  & 8.71\%  \\
{\tt Asian }          & 3.96\%  & 3.01\%  & 1.50\% & 16.27\% & 1.91\%  & 11.96\% & 0.00\% & 0.55\% & 2.39\% & 16.54\% & 2.67\% & 0.89\% & 7.04\%  & 1.16\% & 1.91\% & 20.16\% & 8.07\%  \\
{\tt White}           & 8.37\%  & 6.93\%  & 1.15\% & 15.74\% & 4.27\%  & 13.56\% & 0.05\% & 1.44\% & 1.07\% & 14.32\% & 2.05\% & 2.24\% & 3.54\%  & 1.39\% & 1.15\% & 13.78\% & 8.95\%  \\
{\tt Black }          & 5.45\%  & 4.05\%  & 0.82\% & 15.00\% & 1.18\%  & 12.45\% & 0.00\% & 0.27\% & 0.55\% & 28.55\% & 2.18\% & 0.64\% & 8.45\%  & 0.82\% & 1.27\% & 12.36\% & 5.91\%  \\
{\tt Hispanic  }      & 7.55\%  & 6.60\%  & 1.89\% & 11.32\% & 3.77\%  & 7.55\%  & 0.00\% & 0.00\% & 0.00\% & 23.58\% & 3.77\% & 1.89\% & 10.38\% & 0.94\% & 2.83\% & 11.32\% & 6.60\%  \\
Following Trump & 8.50\%  & 5.88\%  & 0.44\% & 12.85\% & 11.33\% & 10.68\% & 0.00\% & 2.61\% & 1.09\% & 16.34\% & 2.40\% & 5.66\% & 2.83\%  & 0.22\% & 2.61\% & 12.42\% & 4.14\%  \\
Following Biden & 13.44\% & 9.68\%  & 0.72\% & 19.62\% & 0.72\%  & 18.01\% & 0.00\% & 0.27\% & 0.45\% & 12.46\% & 1.08\% & 0.18\% & 1.79\%  & 0.81\% & 0.90\% & 9.68\%  & 10.22\%\\
\bottomrule
\end{tabular}
    \end{threeparttable}
\end{sidewaystable*}

\begin{table*}[htbp!]
\centering
\begin{threeparttable}
\caption{Level 2 topic distribution of ``Double standard'' by user characteristics.}
    \label{tab:level2_topic_double}
    \centering

\begin{tabular}{llll}
\toprule
                & DB.1     & DB.2   & DB.3   \\
                \midrule
\textbf{Total}           & 95.88\%  & 3.35\% & 0.78\% \\
Male            & 95.16\%  & 4.09\% & 0.75\% \\
Female          & 96.65\%  & 2.54\% & 0.81\% \\
Age (18-29)     & 96.64\%  & 3.26\% & 0.10\% \\
Age (30-49)     & 95.22\%  & 3.73\% & 1.05\% \\
Age (50-64)     & 93.92\%  & 3.31\% & 2.76\% \\
Age (65+)       & 96.18\%  & 0.76\% & 3.05\% \\
{\tt Asian}           & 97.26\%  & 2.06\% & 0.69\% \\
{\tt White }          & 97.43\%  & 1.26\% & 1.32\% \\
{\tt Black  }         & 92.93\%  & 6.92\% & 0.15\% \\
{\tt Hispanic}        & 100.00\% & 0.00\% & 0.00\% \\
Following Trump & 95.58\%  & 1.20\% & 3.21\% \\
Following Biden & 96.25\%  & 3.33\% & 0.42\%\\
\bottomrule
\end{tabular}
    \end{threeparttable}
\end{table*}

\begin{table*}[htbp!]
\centering
\begin{threeparttable}
\caption{Level 2 topic distribution of ``Negative opinion'' by user characteristics.}
    \label{tab:level2_topic_negative}
    \centering

\begin{tabular}{lllllll}
\toprule
                & NO.1    & NO.2   & NO.3   & NO.4   & NO.5   & NO.6    \\
                \midrule
\textbf{Total}           & 93.73\% & 0.72\% & 1.00\% & 1.04\% & 0.40\% & 3.12\%  \\
Male            & 91.70\% & 1.21\% & 1.05\% & 1.45\% & 0.32\% & 4.27\%  \\
Female          & 95.72\% & 0.24\% & 0.95\% & 0.63\% & 0.48\% & 1.98\%  \\
Age (18-29)     & 95.84\% & 0.67\% & 0.13\% & 1.21\% & 0.40\% & 1.74\%  \\
Age (30-49)     & 91.05\% & 0.90\% & 1.53\% & 0.90\% & 0.38\% & 5.24\%  \\
Age (50-64)     & 87.64\% & 0.56\% & 5.62\% & 0.56\% & 0.56\% & 5.06\%  \\
Age (65+)       & 94.00\% & 0.00\% & 2.00\% & 0.00\% & 0.00\% & 4.00\%  \\
{\tt Asian}           & 94.97\% & 0.44\% & 0.66\% & 0.44\% & 0.00\% & 3.50\%  \\
{\tt White}           & 89.43\% & 0.38\% & 2.64\% & 0.63\% & 0.25\% & 6.67\%  \\
{\tt Black }          & 96.01\% & 1.06\% & 0.08\% & 1.47\% & 0.65\% & 0.73\%  \\
{\tt Hispanic}        & 95.65\% & 0.00\% & 0.00\% & 4.35\% & 0.00\% & 0.00\%  \\
Following Trump & 92.78\% & 0.00\% & 2.06\% & 1.03\% & 0.52\% & 3.61\%  \\
Following Biden & 81.48\% & 0.00\% & 1.23\% & 0.00\% & 0.00\% & 17.28\%\\
\bottomrule
\end{tabular}

    \end{threeparttable}
\end{table*}

\begin{table*}[htbp!]
\centering
\begin{threeparttable}
\caption{Level 2 topic distribution of ``Policy'' by user characteristics.}
    \label{tab:level2_topic_policy}
    \centering

\begin{tabular}{lll}
\toprule
                & P.1     & P.2     \\
                \midrule
\textbf{Total}           & 87.62\% & 12.38\% \\
Male            & 88.17\% & 11.83\% \\
Female          & 87.27\% & 12.73\% \\
Age (18-29)     & 88.76\% & 11.24\% \\
Age (30-49)     & 88.69\% & 11.31\% \\
Age (50-64)     & 80.99\% & 19.01\% \\
Age (65+)       & 91.67\% & 8.33\%  \\
{\tt Asian      }     & 89.11\% & 10.89\% \\
{\tt White     }      & 89.13\% & 10.87\% \\
{\tt Black    }       & 76.67\% & 23.33\% \\
{\tt Hispanic}        & 92.31\% & 7.69\%  \\
Following Trump & 94.74\% & 5.26\%  \\
Following Biden & 86.41\% & 13.59\%\\
\bottomrule
\end{tabular}

    \end{threeparttable}
\end{table*}

\subsection{Logistic regression outputs}
In this section, we show the logistic regression outputs for level 1 topics (Table~\ref{tab:level1_logit_full}) and level 2 topics including ``Support'' (Table~\ref{tab:level2_logit_support}), ``News'' (Table~\ref{tab:level2_logit_news}), ``Denouncement'' (Table~\ref{tab:level2_logit_denouncement}), ``Double standard'' (Table~\ref{tab:level2_logit_double}), and ``Policy'' (Table~\ref{tab:level2_logit_policy}). We do not conduct logistic regression for level 2 topics of ``Negative opinion'' due to the small sample size. Control variables include social capital of the users, income, religious status, family status, and population density.

\begin{table*}[htbp!]
\small
\centering
\begin{threeparttable}
\caption{Full logistic regression outputs for the opinion on the \#StopAsianHate and \#StopAAPIHate movement against demographics and other variables of interest.}
    \label{tab:level1_logit_full}
    \centering
    
    \begin{tabular}{l l l l l l l}
    \toprule
    Independent variable     & Support & News & Denouncement & Double standard & Negative opinion & Policy \\
     & (1) & (2) & (3) & (4) & (5) & (6)\\
         \midrule
        Male & \makecell[l]{-$0.27^{***}$ \\ (0.02)} & \makecell[l]{$0.30^{***}$ \\ (0.03)} & \makecell[l]{0.01 \\ (0.03)} & \makecell[l]{$0.28^{***}$ \\ (0.04)} & \makecell[l]{$0.14^{**}$ \\ (0.04)}& \makecell[l]{-$0.23^{**}$\\(0.08)  }\\
        
        Age (years) & \makecell[l]{-$0.01^{***}$ \\ (0.00)} &  \makecell[l]{$0.01^{***}$ \\ (0.00)} & \makecell[l]{$0.01^{***}$ \\ (0.00)} & \makecell[l]{-0.00 \\ (0.00)} & \makecell[l]{-$0.01^{***}$ \\ (0.00)} & \makecell[l]{$0.01^{**}$\\(0.00)}\\
        
        {\tt Verified    }         & \begin{tabular}[c]{@{}l@{}}-$0.20^{**}$\\ (0.07)\end{tabular}      & \begin{tabular}[c]{@{}l@{}}$0.27^{**}$\\ (0.08)\end{tabular}       & \begin{tabular}[c]{@{}l@{}}-0.19\\ (0.10)\end{tabular}             & \begin{tabular}[c]{@{}l@{}}$0.66^{***}$\\ (0.14)\end{tabular}       & \begin{tabular}[c]{@{}l@{}}-0.51\\ (0.32)\end{tabular}         & \begin{tabular}[c]{@{}l@{}}-$0.91^{**}$\\ (0.31)\end{tabular}  \\
{\tt Followers   }         & \begin{tabular}[c]{@{}l@{}}-$0.07^{***}$\\ (0.01)\end{tabular}     & \begin{tabular}[c]{@{}l@{}}-0.01\\ (0.02)\end{tabular}             & \begin{tabular}[c]{@{}l@{}}-$0.05^{**}$\\ (0.02)\end{tabular}      & \begin{tabular}[c]{@{}l@{}}$0.08^{***}$\\ (0.02)\end{tabular}       & \begin{tabular}[c]{@{}l@{}}$0.31^{***}$\\ (0.03)\end{tabular}  & \begin{tabular}[c]{@{}l@{}}-0.08\\ (0.05)\end{tabular}         \\
{\tt Friends }             & \begin{tabular}[c]{@{}l@{}}$0.04^{**}$\\ (0.01)\end{tabular}       & \begin{tabular}[c]{@{}l@{}}0.02\\ (0.02)\end{tabular}              & \begin{tabular}[c]{@{}l@{}}$0.05^{**}$\\ (0.02)\end{tabular}       & \begin{tabular}[c]{@{}l@{}}-$0.14^{***}$\\ (0.02)\end{tabular}      & \begin{tabular}[c]{@{}l@{}}-$0.16^{***}$\\ (0.03)\end{tabular} & \begin{tabular}[c]{@{}l@{}}0.02\\ (0.05)\end{tabular}          \\
{\tt List memberships}     & \begin{tabular}[c]{@{}l@{}}$0.17^{***}$\\ (0.02)\end{tabular}      & \begin{tabular}[c]{@{}l@{}}$0.08^{***}$\\ (0.02)\end{tabular}      & \begin{tabular}[c]{@{}l@{}}0.04\\ (0.02)\end{tabular}              & \begin{tabular}[c]{@{}l@{}}-$0.40^{***}$\\ (0.03)\end{tabular}      & \begin{tabular}[c]{@{}l@{}}-$0.08^{***}$\\ (0.02)\end{tabular} & \begin{tabular}[c]{@{}l@{}}$0.31^{**}$\\ (0.06)\end{tabular}   \\
{\tt Favourites }          & \begin{tabular}[c]{@{}l@{}}$0.06^{***}$\\ (0.01)\end{tabular}      & \begin{tabular}[c]{@{}l@{}}-$0.06^{***}$\\ (0.01)\end{tabular}     & \begin{tabular}[c]{@{}l@{}}-0.00\\ (0.01)\end{tabular}             & \begin{tabular}[c]{@{}l@{}}0.02\\ (0.01)\end{tabular}               & \begin{tabular}[c]{@{}l@{}}-$0.08^{***}$\\ (0.02)\end{tabular} & \begin{tabular}[c]{@{}l@{}}$0.08^{*}$\\ (0.03)\end{tabular}    \\
{\tt Statuses }            & \begin{tabular}[c]{@{}l@{}}-$0.14^{***}$\\ (0.01)\end{tabular}     & \begin{tabular}[c]{@{}l@{}}$0.06^{***}$\\ (0.01)\end{tabular}      & \begin{tabular}[c]{@{}l@{}}-$0.03^{*}$\\ (0.01)\end{tabular}       & \begin{tabular}[c]{@{}l@{}}$0.27^{***}$\\ (0.02)\end{tabular}       & \begin{tabular}[c]{@{}l@{}}$0.22^{***}$\\ (0.02)\end{tabular}  & \begin{tabular}[c]{@{}l@{}}-$0.18^{***}$\\ (0.03)\end{tabular} \\
Income               & \begin{tabular}[c]{@{}l@{}}-$2.76e-6^{*}$\\ (1.11e-6)\end{tabular} & \begin{tabular}[c]{@{}l@{}}$4.00e-6^{**}$\\ (1.49e-6)\end{tabular} & \begin{tabular}[c]{@{}l@{}}$4.20e-6^{**}$\\ (1.49e-6)\end{tabular} & \begin{tabular}[c]{@{}l@{}}-$6.88e-6^{**}$\\ (2.17e-6)\end{tabular} & \begin{tabular}[c]{@{}l@{}}-4.15e-6\\ (2.73e-6)\end{tabular}   & \begin{tabular}[c]{@{}l@{}}3.40e-6\\ (4.16e-6)\end{tabular}    \\
Religious            & \begin{tabular}[c]{@{}l@{}}-$0.33^{***}$\\ (0.04)\end{tabular}     & \begin{tabular}[c]{@{}l@{}}$1.24^{***}$\\ (0.05)\end{tabular}      & \begin{tabular}[c]{@{}l@{}}-$0.78^{***}$\\ (0.07)\end{tabular}     & \begin{tabular}[c]{@{}l@{}}-$0.65^{***}$\\ (0.09)\end{tabular}      & \begin{tabular}[c]{@{}l@{}}-$0.78^{***}$\\ (0.12)\end{tabular} & \begin{tabular}[c]{@{}l@{}}-$1.45^{***}$\\ (0.29)\end{tabular} \\
Family               & \begin{tabular}[c]{@{}l@{}}-0.05\\ (0.04)\end{tabular}             & \begin{tabular}[c]{@{}l@{}}0.04\\ (0.05)\end{tabular}              & \begin{tabular}[c]{@{}l@{}}$0.13^{**}$\\ (0.05)\end{tabular}       & \begin{tabular}[c]{@{}l@{}}-0.10\\ (0.08)\end{tabular}              & \begin{tabular}[c]{@{}l@{}}-$0.27^{*}$\\ (0.12)\end{tabular}   & \begin{tabular}[c]{@{}l@{}}0.20\\ (0.13)\end{tabular}          \\

        Following Trump & \makecell[l]{-$0.59^{***}$ \\ (0.04)} &  \makecell[l]{$0.51^{***}$ \\ (0.05)} & \makecell[l]{0.09 \\ (0.05)} & \makecell[l]{$0.22^{**}$ \\ (0.07)} & \makecell[l]{$0.61^{***}$ \\ (0.08)} & \makecell[l]{-0.19 \\ (0.17)}\\
        
        Following Biden & \makecell[l]{$0.24^{***}$ \\ (0.03)} &  \makecell[l]{-0.01 \\ (0.04)} & \makecell[l]{0.07 \\ (0.04)} & \makecell[l]{-$0.81^{***}$ \\ (0.07)} & \makecell[l]{-$1.26^{***}$ \\ (0.12)} & \makecell[l]{$0.56^{***}$ \\ (0.09)}\\
        
        {\tt White} & \makecell[l]{-$0.25^{***}$ \\ (0.02)} &  \makecell[l]{0.06 \\ (0.03)} & \makecell[l]{$0.27^{***}$ \\ (0.03)} & \makecell[l]{$0.17^{***}$ \\ (0.05)} & \makecell[l]{-0.06 \\ (0.06)} & \makecell[l]{$0.84^{***}$ \\ (0.11)} \\
        
        {\tt Black} & \makecell[l]{-$0.46^{***}$ \\ (0.03)} &  \makecell[l]{-$0.34^{***}$ \\ (0.04)} & \makecell[l]{0.02 \\ (0.04)} & \makecell[l]{$0.72^{***}$ \\ (0.05)} & \makecell[l]{$1.10^{***}$ \\ (0.06)} & \makecell[l]{$0.36^{*}$ \\ (0.15)}\\
        
        {\tt Hispanic} & \makecell[l]{-0.12 \\ (0.08)} &  \makecell[l]{0.01 \\ (0.10)} & \makecell[l]{$0.22^{*}$ \\ (0.11)} & \makecell[l]{0.02 \\ (0.16)} & \makecell[l]{-0.23 \\ (0.22)} & \makecell[l]{$0.84^{**}$ \\ (0.30)}\\

Urban                & \begin{tabular}[c]{@{}l@{}}-0.01\\ (0.03)\end{tabular}             & \begin{tabular}[c]{@{}l@{}}0.06\\ (0.04)\end{tabular}              & \begin{tabular}[c]{@{}l@{}}-0.04\\ (0.04)\end{tabular}             & \begin{tabular}[c]{@{}l@{}}-0.02\\ (0.05)\end{tabular}              & \begin{tabular}[c]{@{}l@{}}0.05\\ (0.07)\end{tabular}          & \begin{tabular}[c]{@{}l@{}}0.08\\ ()0.11\end{tabular}          \\
Suburban             & \begin{tabular}[c]{@{}l@{}}0.01\\ (0.04)\end{tabular}              & \begin{tabular}[c]{@{}l@{}}0.04\\ (0.05)\end{tabular}              & \begin{tabular}[c]{@{}l@{}}-0.09\\ (0.05)\end{tabular}             & \begin{tabular}[c]{@{}l@{}}-0.03\\ (0.07)\end{tabular}              & \begin{tabular}[c]{@{}l@{}}0.13\\ (0.08)\end{tabular}          & \begin{tabular}[c]{@{}l@{}}0.01\\ (0.15)\end{tabular}          \\

        Hate crimes & \makecell[l]{0.16 \\ (0.10)} &  \makecell[l]{-0.02 \\ (0.13)} & \makecell[l]{0.23 \\ (0.14)} & \makecell[l]{-0.06 \\ (0.19)} & \makecell[l]{-$1.80^{***}$ \\ (0.27)}&\makecell[l]{0.22 \\ (0.38)}\\

        N &  \multicolumn{6}{c}{46,085}\\
    \bottomrule
    
    \end{tabular}
    \begin{tablenotes}
    \small
    \item Note. * $p<0.05$. ** $p<0.01$. *** $p<0.001$. Table entries are coefficients (standard errors).
    \end{tablenotes}
    \end{threeparttable}
\end{table*}

\begin{table*}[htbp!]
\small
\centering
\begin{threeparttable}
\caption{Logistic regression outputs for the level 2 topics of ``Support'' on the \#StopAsianHate and \#StopAAPIHate movement against demographics and other variables of interest.}
    \label{tab:level2_logit_support}
    \centering
    
    \begin{tabular}{l l l l l l l l l l}
    \toprule
    Independent variable     & S.1 & S.2 & S.3 & S.4 & S.5 & S.6 & S.7 & S.8 & S.9\\

         \midrule
        Male & \makecell[l]{$0.16^{***}$ \\ (0.03)} & \makecell[l]{-$0.54^{***}$ \\ (0.04)} & \makecell[l]{-$0.50^{***}$ \\ (0.10)} & \makecell[l]{$0.20^{**}$ \\ (0.06)} & \makecell[l]{$0.38^{***}$ \\ (0.05)}& \makecell[l]{-0.01\\(0.06)  }& \makecell[l]{-0.16\\(0.09)  }& \makecell[l]{-$0.34^{***}$\\(0.07)  }& \makecell[l]{$0.99^{***}$\\(0.15)}\\
        
        Age (years) & \makecell[l]{$0.01^{***}$ \\ (0.00)} &  \makecell[l]{-$0.03^{***}$ \\ (0.00)} & \makecell[l]{$0.01^{**}$ \\ (0.00)} & \makecell[l]{$0.03^{***}$ \\ (0.00)} & \makecell[l]{-$0.03^{***}$ \\ (0.00)} & \makecell[l]{-$0.03^{***}$\\(0.00)}& \makecell[l]{-$0.02^{***}$\\(0.00)  }& \makecell[l]{$0.01^{***}$\\(0.00)}& \makecell[l]{$0.01^{***}$\\(0.00)}\\
        
       {\tt  Verified   }          & \begin{tabular}[c]{@{}l@{}}$0.26^{**}$\\ (0.10)\end{tabular}   & \begin{tabular}[c]{@{}l@{}}-$0.59^{***}$\\ (0.14)\end{tabular} & \begin{tabular}[c]{@{}l@{}}0.05\\ (0.27)\end{tabular}          & \begin{tabular}[c]{@{}l@{}}0.19\\ (0.19)\end{tabular}          & \begin{tabular}[c]{@{}l@{}}0.19\\ (0.19)\end{tabular}          & \begin{tabular}[c]{@{}l@{}}-0.49\\ (0.28)\end{tabular}         & \begin{tabular}[c]{@{}l@{}}0.04\\ (0.29)\end{tabular}         & \begin{tabular}[c]{@{}l@{}}-0.16\\ (0.36)\end{tabular}         & \begin{tabular}[c]{@{}l@{}}-0.18\\ (0.76)\end{tabular}         \\
{\tt Followers }           & \begin{tabular}[c]{@{}l@{}}-$0.06^{***}$\\ (0.02)\end{tabular} & \begin{tabular}[c]{@{}l@{}}$0.06^{*}$\\ (0.02)\end{tabular}    & \begin{tabular}[c]{@{}l@{}}-0.05\\ (0.06)\end{tabular}         & \begin{tabular}[c]{@{}l@{}}-$0.14^{***}$\\ (0.04)\end{tabular} & \begin{tabular}[c]{@{}l@{}}-$0.07^{*}$\\ (0.03)\end{tabular}   & \begin{tabular}[c]{@{}l@{}}$0.17^{***}$\\ (0.04)\end{tabular}  & \begin{tabular}[c]{@{}l@{}}0.07\\ (0.05)\end{tabular}         & \begin{tabular}[c]{@{}l@{}}$0.20^{***}$\\ (0.04)\end{tabular}  & \begin{tabular}[c]{@{}l@{}}$0.20^{*}$\\ (0.09)\end{tabular}    \\
{\tt Friends    }          & \begin{tabular}[c]{@{}l@{}}$0.04^{*}$\\ (0.02)\end{tabular}    & \begin{tabular}[c]{@{}l@{}}-0.02\\ (0.02)\end{tabular}         & \begin{tabular}[c]{@{}l@{}}-0.07\\ (0.06)\end{tabular}         & \begin{tabular}[c]{@{}l@{}}$0.21^{***}$\\ (0.04)\end{tabular}  & \begin{tabular}[c]{@{}l@{}}0.04\\ (0.03)\end{tabular}          & \begin{tabular}[c]{@{}l@{}}-$0.16^{***}$\\ (0.04)\end{tabular} & \begin{tabular}[c]{@{}l@{}}0.01\\ (0.05)\end{tabular}         & \begin{tabular}[c]{@{}l@{}}-$0.32^{***}$\\ (0.04)\end{tabular} & \begin{tabular}[c]{@{}l@{}}0.08\\ (0.08)\end{tabular}          \\
{\tt List memberships}     & \begin{tabular}[c]{@{}l@{}}$0.07^{**}$\\ (0.02)\end{tabular}   & \begin{tabular}[c]{@{}l@{}}$0.24^{***}$\\ (0.03)\end{tabular}  & \begin{tabular}[c]{@{}l@{}}$0.25^{**}$\\ (0.08)\end{tabular}   & \begin{tabular}[c]{@{}l@{}}0.04\\ (0.05)\end{tabular}          & \begin{tabular}[c]{@{}l@{}}-$0.18^{***}$\\ (0.05)\end{tabular} & \begin{tabular}[c]{@{}l@{}}-$0.16^{**}$\\ (0.06)\end{tabular}  & \begin{tabular}[c]{@{}l@{}}-0.01\\ (0.07)\end{tabular}        & \begin{tabular}[c]{@{}l@{}}-$0.77^{***}$\\ (0.07)\end{tabular} & \begin{tabular}[c]{@{}l@{}}-$0.68^{***}$\\ (0.15)\end{tabular} \\
{\tt Favourites}           & \begin{tabular}[c]{@{}l@{}}-$0.06^{***}$\\ (0.01)\end{tabular} & \begin{tabular}[c]{@{}l@{}}$0.15^{***}$\\ (0.01)\end{tabular}  & \begin{tabular}[c]{@{}l@{}}$0.16^{***}$\\ (0.04)\end{tabular}  & \begin{tabular}[c]{@{}l@{}}-$0.18^{***}$\\ (0.02)\end{tabular} & \begin{tabular}[c]{@{}l@{}}-$0.07^{***}$\\ (0.02)\end{tabular} & \begin{tabular}[c]{@{}l@{}}$0.09^{***}$\\ (0.03)\end{tabular}  & \begin{tabular}[c]{@{}l@{}}$0.07^{*}$\\ (0.03)\end{tabular}   & \begin{tabular}[c]{@{}l@{}}$0.07^{**}$\\ (0.03)\end{tabular}   & \begin{tabular}[c]{@{}l@{}}-$0.15^{**}$\\ (0.05)\end{tabular}  \\
{\tt Statuses}             & \begin{tabular}[c]{@{}l@{}}$0.05^{***}$\\ (0.01)\end{tabular}  & \begin{tabular}[c]{@{}l@{}}-$0.28^{***}$\\ (0.02)\end{tabular} & \begin{tabular}[c]{@{}l@{}}-$0.25^{***}$\\ (0.04)\end{tabular} & \begin{tabular}[c]{@{}l@{}}$0.20^{***}$\\ (0.03)\end{tabular}  & \begin{tabular}[c]{@{}l@{}}$0.29^{***}$\\ (0.02)\end{tabular}  & \begin{tabular}[c]{@{}l@{}}-$0.15^{***}$\\ (0.03)\end{tabular} & \begin{tabular}[c]{@{}l@{}}-0.06\\ (0.04)\end{tabular}        & \begin{tabular}[c]{@{}l@{}}-0.01\\ (0.03)\end{tabular}         & \begin{tabular}[c]{@{}l@{}}0.07\\ (0.06)\end{tabular}          \\
Income               & \begin{tabular}[c]{@{}l@{}}-1.58e-6\\ (1.57e-6)\end{tabular}   & \begin{tabular}[c]{@{}l@{}}2.72e-7\\ (2.19e-6)\end{tabular}    & \begin{tabular}[c]{@{}l@{}}-2.78e-6\\ (5.40e-6)\end{tabular}   & \begin{tabular}[c]{@{}l@{}}5.12e-6\\ (3.21e-6)\end{tabular}    & \begin{tabular}[c]{@{}l@{}}-1.12e-6\\ (2.94e-6)\end{tabular}   & \begin{tabular}[c]{@{}l@{}}-3.46e-6\\ (3.67e-6)\end{tabular}   & \begin{tabular}[c]{@{}l@{}}3.55e-6\\ (4.73e-6)\end{tabular}   & \begin{tabular}[c]{@{}l@{}}5.38e-6\\ (3.29e-6)\end{tabular}    & \begin{tabular}[c]{@{}l@{}}1.23e-5\\ (7.03e-6)\end{tabular}    \\
Religious            & \begin{tabular}[c]{@{}l@{}}-$1.08^{***}$\\ (0.06)\end{tabular} & \begin{tabular}[c]{@{}l@{}}-$1.41^{***}$\\ (0.13)\end{tabular} & \begin{tabular}[c]{@{}l@{}}-$1.26^{***}$\\ (0.36)\end{tabular} & \begin{tabular}[c]{@{}l@{}}-0.10\\ (0.14)\end{tabular}         & \begin{tabular}[c]{@{}l@{}}-$1.34^{***}$\\ (0.17)\end{tabular} & \begin{tabular}[c]{@{}l@{}}$3.36^{***}$\\ (0.07)\end{tabular}  & \begin{tabular}[c]{@{}l@{}}-$0.57^{**}$\\ (0.21)\end{tabular} & \begin{tabular}[c]{@{}l@{}}-$1.33^{***}$\\ (0.22)\end{tabular} & \begin{tabular}[c]{@{}l@{}}$0.55^{*}$\\ (0.24)\end{tabular}    \\
Family               & \begin{tabular}[c]{@{}l@{}}$0.38^{***}$\\ (0.06)\end{tabular}  & \begin{tabular}[c]{@{}l@{}}-$0.41^{***}$\\ (0.09)\end{tabular} & \begin{tabular}[c]{@{}l@{}}-0.17\\ (0.18)\end{tabular}         & \begin{tabular}[c]{@{}l@{}}0.04\\ (0.10)\end{tabular}          & \begin{tabular}[c]{@{}l@{}}-$0.55^{***}$\\ (0.14)\end{tabular} & \begin{tabular}[c]{@{}l@{}}-$0.67^{***}$\\ (0.16)\end{tabular} & \begin{tabular}[c]{@{}l@{}}-0.07\\ (0.19)\end{tabular}        & \begin{tabular}[c]{@{}l@{}}0.07\\ (0.12)\end{tabular}          & \begin{tabular}[c]{@{}l@{}}0.44\\ (0.23)\end{tabular}          \\

         Following Trump & \makecell[l]{0.04 \\ (0.07)} &  \makecell[l]{-$0.45^{***}$ \\ (0.11)} & \makecell[l]{-0.00 \\ (0.22)} & \makecell[l]{$0.29^{*}$ \\ (0.12)} & \makecell[l]{-0.13 \\ (0.13)} & \makecell[l]{-0.22 \\ (0.16)}& \makecell[l]{-0.24\\(0.24)  }& \makecell[l]{-0.07\\(0.15)}& \makecell[l]{$1.82^{***}$\\(0.17)}\\
        
        Following Biden & \makecell[l]{$0.22^{***}$ \\ (0.04)} &  \makecell[l]{-$0.64^{***}$ \\ (0.06)} & \makecell[l]{$0.28^{*}$ \\ (0.12)} & \makecell[l]{$0.89^{***}$ \\ (0.07)} & \makecell[l]{-$0.65^{***}$ \\ (0.09)} & \makecell[l]{-$0.31^{**}$ \\ (0.10)}& \makecell[l]{-$0.71^{***}$\\(0.16)  }& \makecell[l]{$0.20^{*}$\\(0.09)}& \makecell[l]{-$1.85^{***}$\\(0.33)}\\
    Urban                & \begin{tabular}[c]{@{}l@{}}0.00\\ (0.04)\end{tabular}          & \begin{tabular}[c]{@{}l@{}}0.11\\ (0.06)\end{tabular}          & \begin{tabular}[c]{@{}l@{}}$0.52^{**}$\\ (0.15)\end{tabular}   & \begin{tabular}[c]{@{}l@{}}-$0.25^{**}$\\ (0.08)\end{tabular}  & \begin{tabular}[c]{@{}l@{}}$0.24^{**}$\\ (0.07)\end{tabular}   & \begin{tabular}[c]{@{}l@{}}0.06\\ (0.09)\end{tabular}          & \begin{tabular}[c]{@{}l@{}}$0.39^{**}$\\ (0.14)\end{tabular}  & \begin{tabular}[c]{@{}l@{}}-$0.22^{**}$\\ (0.08)\end{tabular}  & \begin{tabular}[c]{@{}l@{}}-$1.05^{***}$\\ (0.16)\end{tabular} \\
Suburban             & \begin{tabular}[c]{@{}l@{}}0.02\\ (0.05)\end{tabular}          & \begin{tabular}[c]{@{}l@{}}0.10\\ (0.08)\end{tabular}          & \begin{tabular}[c]{@{}l@{}}0.20\\ (0.20)\end{tabular}          & \begin{tabular}[c]{@{}l@{}}-0.03\\ (0.11)\end{tabular}         & \begin{tabular}[c]{@{}l@{}}-0.11\\ (0.10)\end{tabular}         & \begin{tabular}[c]{@{}l@{}}-0.10\\ (0.12)\end{tabular}         & \begin{tabular}[c]{@{}l@{}}0.27\\ (0.18)\end{tabular}         & \begin{tabular}[c]{@{}l@{}}0.06\\ (0.11)\end{tabular}          & \begin{tabular}[c]{@{}l@{}}-$0.68^{**}$\\ (0.22)\end{tabular}  \\

        {\tt White} & \makecell[l]{$0.53^{***}$ \\ (0.03)} &  \makecell[l]{$0.38^{***}$ \\ (0.04)} & \makecell[l]{$0.37^{***}$ \\ (0.11)} & \makecell[l]{$0.51^{***}$ \\ (0.08)} & \makecell[l]{-$0.38^{***}$ \\ (0.06)} & \makecell[l]{0.07 \\ (0.07)} & \makecell[l]{-$0.38^{***}$\\(0.10)  }& - & \makecell[l]{$1.48^{***}$\\(0.24)}\\
        
        {\tt Black} & \makecell[l]{$0.64^{***}$ \\ (0.04)} &  \makecell[l]{-$0.32^{***}$ \\ (0.06)} & \makecell[l]{-0.19 \\ (0.17)} & \makecell[l]{0.08 \\ (0.11)} & \makecell[l]{$0.13^{*}$ \\ (0.06)} & \makecell[l]{0.12 \\ (0.09)}& \makecell[l]{$0.89^{***}$\\(0.10)  }& - & \makecell[l]{-0.38\\(0.38)}\\
        
        {\tt Hispanic} & \makecell[l]{$0.64^{***}$ \\ (0.04)} &  \makecell[l]{0.23 \\ (0.14)} & \makecell[l]{-0.51 \\ (0.51)} & \makecell[l]{0.35 \\ (0.26)} & \makecell[l]{0.08 \\ (0.17)} & \makecell[l]{0.16 \\ (0.22)}& \makecell[l]{-0.79\\(0.46)  }& -& \makecell[l]{$1.68^{***}$\\(0.46)}\\
        
        Hate crimes & \makecell[l]{0.20 \\ (0.14)} &  \makecell[l]{$0.60^{***}$ \\ (0.19)} & \makecell[l]{-0.79 \\ (0.48)} & \makecell[l]{-$0.83^{*}$ \\ (0.32)} & \makecell[l]{-$0.87^{**}$ \\ (0.27)}&\makecell[l]{-0.14 \\ (0.32)}& \makecell[l]{0.32\\(0.42)  }& \makecell[l]{0.20\\(0.31)}& \makecell[l]{-0.49\\(0.75)}\\
        
        N &  \multicolumn{9}{c}{23,747}\\
    \bottomrule
    
    \end{tabular}
    \begin{tablenotes}
    \small
    \item Note. * $p<0.05$. ** $p<0.01$. *** $p<0.001$. Table entries are coefficients (standard errors).
    \end{tablenotes}
    \end{threeparttable}
\end{table*}

\begin{table*}[htbp!]

\centering
\begin{threeparttable}
\caption{Logistic regression outputs for the level 2 topics of ``News'' on the \#StopAsianHate and \#StopAAPIHate movement against demographics and other variables of interest.}
    \label{tab:level2_logit_news}
    \centering
\begin{tabular}{lllllll}
\toprule
Independent variable & N.1                                                           & N.2                                                                            & N.3                                                            & N.4                                                            & N.5                                                                 & N.6                                                           \\
\midrule
Male                 & \begin{tabular}[c]{@{}l@{}}$0.28^{***}$\\ (0.05)\end{tabular} & \begin{tabular}[c]{@{}l@{}}$0.19^{*}$\\ (0.09)\end{tabular}   & \begin{tabular}[c]{@{}l@{}}-$0.43^{*}$\\ (0.20)\end{tabular}   & \begin{tabular}[c]{@{}l@{}}-$0.44^{***}$\\ (0.06)\end{tabular} & \begin{tabular}[c]{@{}l@{}}0.12\\ (0.10)\end{tabular}               & \begin{tabular}[c]{@{}l@{}}-0.11\\ (0.29)\end{tabular}        \\
Age (years)                  & \begin{tabular}[c]{@{}l@{}}$0.00^{**}$\\ (0.00)\end{tabular}  & \begin{tabular}[c]{@{}l@{}}$0.03^{***}$\\ (0.00)\end{tabular} & \begin{tabular}[c]{@{}l@{}}-$0.04^{***}$\\ (0.01)\end{tabular} & \begin{tabular}[c]{@{}l@{}}-$0.01^{***}$\\ (0.00)\end{tabular} & \begin{tabular}[c]{@{}l@{}}-$0.02^{***}$\\ (0.00)\end{tabular}      & \begin{tabular}[c]{@{}l@{}}-0.00\\ (0.01)\end{tabular}        \\
{\tt Verified}             & \begin{tabular}[c]{@{}l@{}}0.31\\ (0.16)\end{tabular}         & \begin{tabular}[c]{@{}l@{}}0.03\\ (0.38)\end{tabular}                          & \begin{tabular}[c]{@{}l@{}}-0.91\\ (1.07)\end{tabular}         & \begin{tabular}[c]{@{}l@{}}-$0.55^{**}$\\ (0.20)\end{tabular}  & \begin{tabular}[c]{@{}l@{}}0.22\\ (0.30)\end{tabular}               & \begin{tabular}[c]{@{}l@{}}-0.84\\ (0.87)\end{tabular}        \\
{\tt Followers}            & \begin{tabular}[c]{@{}l@{}}0.00\\ (0.03)\end{tabular}         & \begin{tabular}[c]{@{}l@{}}-0.03\\ (0.05)\end{tabular}                         & \begin{tabular}[c]{@{}l@{}}-0.15\\ (0.12)\end{tabular}         & \begin{tabular}[c]{@{}l@{}}-0.06\\ (0.03)\end{tabular}         & \begin{tabular}[c]{@{}l@{}}$0.19^{**}$\\ (0.06)\end{tabular}        & \begin{tabular}[c]{@{}l@{}}0.07\\ (0.17)\end{tabular}         \\
{\tt Friends}              & \begin{tabular}[c]{@{}l@{}}-0.01\\ (0.03)\end{tabular}        & \begin{tabular}[c]{@{}l@{}}0.09\\ (0.05)\end{tabular}                          & \begin{tabular}[c]{@{}l@{}}0.16\\ (0.11)\end{tabular}          & \begin{tabular}[c]{@{}l@{}}-0.02\\ (0.03)\end{tabular}         & \begin{tabular}[c]{@{}l@{}}-0.01\\ (0.06)\end{tabular}              & \begin{tabular}[c]{@{}l@{}}0.07\\ (0.17)\end{tabular}         \\
{\tt List memberships}     & \begin{tabular}[c]{@{}l@{}}$0.12^{**}$\\ (0.04)\end{tabular}  & \begin{tabular}[c]{@{}l@{}}-$0.35^{***}$\\ (0.08)\end{tabular}                 & \begin{tabular}[c]{@{}l@{}}0.05\\ (0.18)\end{tabular}          & \begin{tabular}[c]{@{}l@{}}0.07\\ (0.05)\end{tabular}          & \begin{tabular}[c]{@{}l@{}}-$0.26^{**}$\\ (0.09)\end{tabular}       & \begin{tabular}[c]{@{}l@{}}0.27\\ (0.23)\end{tabular}         \\
{\tt Favourites}           & \begin{tabular}[c]{@{}l@{}}-$0.06^{**}$\\ (0.02)\end{tabular} & \begin{tabular}[c]{@{}l@{}}-0.05\\ (0.03)\end{tabular}                         & \begin{tabular}[c]{@{}l@{}}0.14\\ (0.08)\end{tabular}          & \begin{tabular}[c]{@{}l@{}}$0.16^{***}$\\ (0.02)\end{tabular}  & \begin{tabular}[c]{@{}l@{}}-$0.14^{***}$\\ (0.03)\end{tabular}      & \begin{tabular}[c]{@{}l@{}}0.07\\ (0.12)\end{tabular}         \\
{\tt Statuses}             & \begin{tabular}[c]{@{}l@{}}$0.05^{*}$\\ (0.02)\end{tabular}   & \begin{tabular}[c]{@{}l@{}}$0.09^{*}$\\ (0.04)\end{tabular}                    & \begin{tabular}[c]{@{}l@{}}-0.10\\ (0.09)\end{tabular}         & \begin{tabular}[c]{@{}l@{}}-$0.15^{***}$\\ (0.03)\end{tabular} & \begin{tabular}[c]{@{}l@{}}$0.16^{***}$\\ (0.04)\end{tabular}       & \begin{tabular}[c]{@{}l@{}}-$0.30^{*}$\\ (0.13)\end{tabular}  \\
Income            & \begin{tabular}[c]{@{}l@{}}6.41e-7\\ (2.64e-6)\end{tabular}   & \begin{tabular}[c]{@{}l@{}}-4.10e-6\\ (4.69e-6)\end{tabular}                   & \begin{tabular}[c]{@{}l@{}}3.57e-6\\ (1.11e-5)\end{tabular}    & \begin{tabular}[c]{@{}l@{}}5.22e-6\\ (3.04e-6)\end{tabular}    & \begin{tabular}[c]{@{}l@{}}-$2.24e-5^{**}$\\ (6.85e-6)\end{tabular} & \begin{tabular}[c]{@{}l@{}}1.58e-5\\ (1.32e-5)\end{tabular}   \\
Religious           & \begin{tabular}[c]{@{}l@{}}$1.86^{***}$\\ (0.10)\end{tabular} & \begin{tabular}[c]{@{}l@{}}-$0.81^{***}$\\ (0.15)\end{tabular}                 & \begin{tabular}[c]{@{}l@{}}-$2.12^{**}$\\ (0.72)\end{tabular}  & \begin{tabular}[c]{@{}l@{}}-$2.10^{***}$\\ (0.15)\end{tabular} & \begin{tabular}[c]{@{}l@{}}-$1.37^{***}$\\ (0.23)\end{tabular}      & -                                                             \\
Family               & \begin{tabular}[c]{@{}l@{}}0.03\\ (0.09)\end{tabular}         & \begin{tabular}[c]{@{}l@{}}0.20\\ (0.14)\end{tabular}                          & \begin{tabular}[c]{@{}l@{}}-0.62\\ (0.52)\end{tabular}         & \begin{tabular}[c]{@{}l@{}}-0.14\\ (0.11)\end{tabular}         & \begin{tabular}[c]{@{}l@{}}-0.32\\ (0.24)\end{tabular}              & \begin{tabular}[c]{@{}l@{}}$1.47^{***}$\\ (0.32)\end{tabular} \\
Following Trump      & \begin{tabular}[c]{@{}l@{}}0.09\\ (0.08)\end{tabular}         & \begin{tabular}[c]{@{}l@{}}$0.94^{***}$\\ (0.11)\end{tabular}                  & \begin{tabular}[c]{@{}l@{}}0.19\\ (0.34)\end{tabular}          & \begin{tabular}[c]{@{}l@{}}-$0.94^{***}$\\ (0.13)\end{tabular} & \begin{tabular}[c]{@{}l@{}}-$0.47^{*}$\\ (0.22)\end{tabular}        & \begin{tabular}[c]{@{}l@{}}0.09\\ (0.48)\end{tabular}         \\
Following Biden      & \begin{tabular}[c]{@{}l@{}}$0.33^{***}$\\ (0.07)\end{tabular} & \begin{tabular}[c]{@{}l@{}}-$1.54^{***}$\\ (0.18)\end{tabular}                 & \begin{tabular}[c]{@{}l@{}}-$0.91^{*}$\\ (0.40)\end{tabular}   & \begin{tabular}[c]{@{}l@{}}$0.24^{**}$\\ (0.08)\end{tabular}   & \begin{tabular}[c]{@{}l@{}}-$0.57^{**}$\\ (0.18)\end{tabular}       & \begin{tabular}[c]{@{}l@{}}0.39\\ (0.35)\end{tabular}         \\
Urban                & \begin{tabular}[c]{@{}l@{}}0.08\\ (0.07)\end{tabular}         & \begin{tabular}[c]{@{}l@{}}-$0.49^{***}$\\ (0.10)\end{tabular}                 & \begin{tabular}[c]{@{}l@{}}0.16\\ (0.31)\end{tabular}          & \begin{tabular}[c]{@{}l@{}}0.16\\ (0.09)\end{tabular}          & \begin{tabular}[c]{@{}l@{}}0.09\\ (0.15)\end{tabular}               & \begin{tabular}[c]{@{}l@{}}-0.33\\ (0.37)\end{tabular}        \\
Suburban             & \begin{tabular}[c]{@{}l@{}}0.07\\ (0.09)\end{tabular}         & \begin{tabular}[c]{@{}l@{}}-0.09\\ (0.13)\end{tabular}                         & \begin{tabular}[c]{@{}l@{}}-0.03\\ (0.41)\end{tabular}         & \begin{tabular}[c]{@{}l@{}}0.09\\ (0.11)\end{tabular}          & \begin{tabular}[c]{@{}l@{}}-0.20\\ (0.21)\end{tabular}              & \begin{tabular}[c]{@{}l@{}}-0.90\\ (0.60)\end{tabular}        \\
{\tt White}                & \begin{tabular}[c]{@{}l@{}}0.05\\ (0.06)\end{tabular}         & \begin{tabular}[c]{@{}l@{}}$0.62^{***}$\\ (0.12)\end{tabular}                  & \begin{tabular}[c]{@{}l@{}}$0.76^{**}$\\ (0.27)\end{tabular}   & \begin{tabular}[c]{@{}l@{}}-$0.28^{***}$\\ (0.07)\end{tabular} & \begin{tabular}[c]{@{}l@{}}-0.11\\ (0.13)\end{tabular}              & \begin{tabular}[c]{@{}l@{}}-0.04\\ (0.35)\end{tabular}        \\
{\tt Black}                & \begin{tabular}[c]{@{}l@{}}-0.13\\ (0.08)\end{tabular}        & \begin{tabular}[c]{@{}l@{}}0.08\\ (0.17)\end{tabular}                          & \begin{tabular}[c]{@{}l@{}}0.58\\ (0.35)\end{tabular}          & \begin{tabular}[c]{@{}l@{}}-$0.33^{***}$\\ (0.09)\end{tabular} & \begin{tabular}[c]{@{}l@{}}$1.02^{***}$\\ (0.14)\end{tabular}       & \begin{tabular}[c]{@{}l@{}}-0.74\\ (0.65)\end{tabular}        \\
{\tt Hispanic}             & \begin{tabular}[c]{@{}l@{}}-0.27\\ (0.19)\end{tabular}        & \begin{tabular}[c]{@{}l@{}}$1.26^{***}$\\ (0.29)\end{tabular}                  & \begin{tabular}[c]{@{}l@{}}$1.78^{***}$\\ (0.49)\end{tabular}  & \begin{tabular}[c]{@{}l@{}}-0.42\\ (0.24)\end{tabular}         & \begin{tabular}[c]{@{}l@{}}-0.10\\ (0.44)\end{tabular}              & \begin{tabular}[c]{@{}l@{}}0.93\\ (0.78)\end{tabular}         \\
Hate crimes          & \begin{tabular}[c]{@{}l@{}}-0.45\\ (0.25)\end{tabular}        & \begin{tabular}[c]{@{}l@{}}-0.16\\ (0.45)\end{tabular}                         & \begin{tabular}[c]{@{}l@{}}1.09\\ (0.98)\end{tabular}          & \begin{tabular}[c]{@{}l@{}}$0.81^{**}$\\ (0.29)\end{tabular}   & \begin{tabular}[c]{@{}l@{}}-0.83\\ (0.59)\end{tabular}              & \begin{tabular}[c]{@{}l@{}}-0.09\\ (1.44)\end{tabular}        \\
N                    & \multicolumn{6}{c}{8,466}     \\                                                 \bottomrule                                                                                                                                                          
\end{tabular}
\begin{tablenotes}
    \small
    \item Note. * $p<0.05$. ** $p<0.01$. *** $p<0.001$. Table entries are coefficients (standard errors).
    \end{tablenotes}
    \end{threeparttable}
\end{table*}

\begin{sidewaystable*}[htbp!]

\centering
\tiny
\begin{threeparttable}
\caption{Logistic regression outputs for the level 2 topics of ``Denouncement'' on the \#StopAsianHate and \#StopAAPIHate movement against demographics and other variables of interest.}
    \label{tab:level2_logit_denouncement}
    \centering
\begin{tabular}{lllllllllllllllll}
\toprule
Independent variable & D.1                                                           & D.2                                                           & D.3                                                          & D.4                                                          & D.5                                                            & D.6                                                            & D.8                                                           & D.9                                                           & D.10                                                           & D.11                                                           & D.12                                                           & D.13                                                           & D.14                                                         & D.15                                                                & D.16                                                           & D.17                                                           \\
\midrule
Male                 & \begin{tabular}[c]{@{}l@{}}0.16\\ (0.10)\end{tabular}         & \begin{tabular}[c]{@{}l@{}}0.08\\ (0.11)\end{tabular}         & \begin{tabular}[c]{@{}l@{}}0.17\\ (0.23)\end{tabular}        & \begin{tabular}[c]{@{}l@{}}-0.08\\ (0.07)\end{tabular}       & \begin{tabular}[c]{@{}l@{}}0.20\\ (0.15)\end{tabular}          & \begin{tabular}[c]{@{}l@{}}$0.50^{***}$\\ (0.08)\end{tabular}  & \begin{tabular}[c]{@{}l@{}}$1.13^{***}$\\ (0.29)\end{tabular} & \begin{tabular}[c]{@{}l@{}}-$0.89^{**}$\\ (0.27)\end{tabular} & \begin{tabular}[c]{@{}l@{}}-0.12\\ (0.07)\end{tabular}         & \begin{tabular}[c]{@{}l@{}}-$0.53^{**}$\\ (0.18)\end{tabular}  & \begin{tabular}[c]{@{}l@{}}$0.80^{***}$\\ (0.22)\end{tabular}  & \begin{tabular}[c]{@{}l@{}}-$0.65^{***}$\\ (0.13)\end{tabular} & \begin{tabular}[c]{@{}l@{}}-$0.54^{*}$\\ (0.24)\end{tabular} & \begin{tabular}[c]{@{}l@{}}$0.60^{**}$\\ (0.22)\end{tabular}        & \begin{tabular}[c]{@{}l@{}}$0.15^{*}$\\ (0.07)\end{tabular}    & \begin{tabular}[c]{@{}l@{}}-$0.52^{***}$\\ (0.10)\end{tabular} \\
Age (years)                 & \begin{tabular}[c]{@{}l@{}}$0.03^{**}$\\ (0.00)\end{tabular}  & \begin{tabular}[c]{@{}l@{}}$0.02^{***}$\\ (0.00)\end{tabular} & \begin{tabular}[c]{@{}l@{}}-0.01\\ (0.01)\end{tabular}       & \begin{tabular}[c]{@{}l@{}}0.00\\ (0.00)\end{tabular}        & \begin{tabular}[c]{@{}l@{}}$0.05^{***}$\\ (0.01)\end{tabular}  & \begin{tabular}[c]{@{}l@{}}0.00\\ (0.00)\end{tabular}          & \begin{tabular}[c]{@{}l@{}}$0.02^{*}$\\ (0.01)\end{tabular}   & \begin{tabular}[c]{@{}l@{}}-$0.03^{**}$\\ (0.01)\end{tabular} & \begin{tabular}[c]{@{}l@{}}-$0.01^{***}$\\ (0.00)\end{tabular} & \begin{tabular}[c]{@{}l@{}}-$0.03^{***}$\\ (0.01)\end{tabular} & \begin{tabular}[c]{@{}l@{}}$0.02^{***}$\\ (0.01)\end{tabular}  & \begin{tabular}[c]{@{}l@{}}-$0.05^{***}$\\ (0.01)\end{tabular} & \begin{tabular}[c]{@{}l@{}}-$0.02^{*}$\\ (0.01)\end{tabular} & \begin{tabular}[c]{@{}l@{}}-$0.02^{*}$\\ (0.01)\end{tabular}        & \begin{tabular}[c]{@{}l@{}}-$0.02^{***}$\\ (0.00)\end{tabular} & \begin{tabular}[c]{@{}l@{}}-0.00\\ (0.00)\end{tabular}         \\
{\tt Verified    }         & \begin{tabular}[c]{@{}l@{}}0.23\\ (0.47)\end{tabular}         & \begin{tabular}[c]{@{}l@{}}-0.31\\ (0.50)\end{tabular}        & -                                                            & \begin{tabular}[c]{@{}l@{}}0.32\\ (0.23)\end{tabular}        & \begin{tabular}[c]{@{}l@{}}-0.00\\ (0.60)\end{tabular}         & \begin{tabular}[c]{@{}l@{}}0.18\\ (0.24)\end{tabular}          & -                                                             & -                                                             & \begin{tabular}[c]{@{}l@{}}-$1.34^{*}$\\ (0.52)\end{tabular}   & \begin{tabular}[c]{@{}l@{}}-1.02\\ (1.06)\end{tabular}         & -                                                              & \begin{tabular}[c]{@{}l@{}}-1.30\\ (1.04)\end{tabular}         & \begin{tabular}[c]{@{}l@{}}0.44\\ (0.52)\end{tabular}        & \begin{tabular}[c]{@{}l@{}}0.99\\ (0.59)\end{tabular}               & \begin{tabular}[c]{@{}l@{}}-0.19\\ (0.24)\end{tabular}         & \begin{tabular}[c]{@{}l@{}}-0.23\\ (0.39)\end{tabular}         \\
{\tt Followers }           & \begin{tabular}[c]{@{}l@{}}-0.03\\ (0.07)\end{tabular}        & \begin{tabular}[c]{@{}l@{}}-0.09\\ (0.07)\end{tabular}        & \begin{tabular}[c]{@{}l@{}}-0.09\\ (0.15)\end{tabular}       & \begin{tabular}[c]{@{}l@{}}-0.04\\ (0.04)\end{tabular}       & \begin{tabular}[c]{@{}l@{}}$0.26^{**}$\\ (0.09)\end{tabular}   & \begin{tabular}[c]{@{}l@{}}-$0.23^{***}$\\ (0.05)\end{tabular} & \begin{tabular}[c]{@{}l@{}}-$0.40^{*}$\\ (0.17)\end{tabular}  & \begin{tabular}[c]{@{}l@{}}0.12\\ (0.14)\end{tabular}         & \begin{tabular}[c]{@{}l@{}}0.08\\ (0.04)\end{tabular}          & \begin{tabular}[c]{@{}l@{}}0.01\\ (0.11)\end{tabular}          & \begin{tabular}[c]{@{}l@{}}0.07\\ (0.13)\end{tabular}          & \begin{tabular}[c]{@{}l@{}}$0.39^{***}$\\ (0.08)\end{tabular}  & \begin{tabular}[c]{@{}l@{}}-0.01\\ (0.14)\end{tabular}       & \begin{tabular}[c]{@{}l@{}}-0.17\\ (0.13)\end{tabular}              & \begin{tabular}[c]{@{}l@{}}0.06\\ (0.04)\end{tabular}          & \begin{tabular}[c]{@{}l@{}}-$0.19^{**}$\\ (0.06)\end{tabular}  \\
{\tt Friends}              & \begin{tabular}[c]{@{}l@{}}0.11\\ (0.06)\end{tabular}         & \begin{tabular}[c]{@{}l@{}}0.09\\ (0.07)\end{tabular}         & \begin{tabular}[c]{@{}l@{}}0.02\\ (0.14)\end{tabular}        & \begin{tabular}[c]{@{}l@{}}0.05\\ (0.04)\end{tabular}        & \begin{tabular}[c]{@{}l@{}}-0.09\\ (0.09)\end{tabular}         & \begin{tabular}[c]{@{}l@{}}$0.19^{***}$\\ (0.05)\end{tabular}  & \begin{tabular}[c]{@{}l@{}}0.07\\ (0.15)\end{tabular}         & \begin{tabular}[c]{@{}l@{}}-0.23\\ (0.14)\end{tabular}        & \begin{tabular}[c]{@{}l@{}}-$0.10^{*}$\\ (0.04)\end{tabular}   & \begin{tabular}[c]{@{}l@{}}-0.09\\ (0.11)\end{tabular}         & \begin{tabular}[c]{@{}l@{}}-0.14\\ (0.12)\end{tabular}         & \begin{tabular}[c]{@{}l@{}}-$0.43^{***}$\\ (0.08)\end{tabular} & \begin{tabular}[c]{@{}l@{}}0.05\\ (0.13)\end{tabular}        & \begin{tabular}[c]{@{}l@{}}0.11\\ (0.13)\end{tabular}               & \begin{tabular}[c]{@{}l@{}}$0.08^{*}$\\ (0.04)\end{tabular}    & \begin{tabular}[c]{@{}l@{}}-0.02\\ (0.06)\end{tabular}         \\
{\tt List memberships}     & \begin{tabular}[c]{@{}l@{}}-$0.33^{**}$\\ (0.10)\end{tabular} & \begin{tabular}[c]{@{}l@{}}0.02\\ (0.10)\end{tabular}         & \begin{tabular}[c]{@{}l@{}}-0.35\\ (0.26)\end{tabular}       & \begin{tabular}[c]{@{}l@{}}0.10\\ (0.06)\end{tabular}        & \begin{tabular}[c]{@{}l@{}}-$0.30^{*}$\\ (0.12)\end{tabular}   & \begin{tabular}[c]{@{}l@{}}$0.40^{***}$\\ (0.06)\end{tabular}  & \begin{tabular}[c]{@{}l@{}}0.11\\ (0.26)\end{tabular}         & \begin{tabular}[c]{@{}l@{}}0.23\\ (0.20)\end{tabular}         & \begin{tabular}[c]{@{}l@{}}-$0.42^{***}$\\ (0.07)\end{tabular} & \begin{tabular}[c]{@{}l@{}}0.03\\ (0.16)\end{tabular}          & \begin{tabular}[c]{@{}l@{}}-$0.58^{**}$\\ (0.21)\end{tabular}  & \begin{tabular}[c]{@{}l@{}}-$0.79^{***}$\\ (0.04)\end{tabular} & \begin{tabular}[c]{@{}l@{}}$0.49^{**}$\\ (0.18)\end{tabular} & \begin{tabular}[c]{@{}l@{}}0.27\\ (0.19)\end{tabular}               & \begin{tabular}[c]{@{}l@{}}$0.28^{***}$\\ (0.06)\end{tabular}  & \begin{tabular}[c]{@{}l@{}}$0.21^{*}$\\ (0.09)\end{tabular}    \\
{\tt Favourites }          & \begin{tabular}[c]{@{}l@{}}-0.03\\ (0.04)\end{tabular}        & \begin{tabular}[c]{@{}l@{}}0.00\\ (0.04)\end{tabular}         & \begin{tabular}[c]{@{}l@{}}$0.23^{*}$\\ (0.10)\end{tabular}  & \begin{tabular}[c]{@{}l@{}}-0.03\\ (0.03)\end{tabular}       & \begin{tabular}[c]{@{}l@{}}-$0.16^{**}$\\ (0.05)\end{tabular}  & \begin{tabular}[c]{@{}l@{}}0.03\\ (0.03)\end{tabular}          & \begin{tabular}[c]{@{}l@{}}-0.01\\ (0.10)\end{tabular}        & \begin{tabular}[c]{@{}l@{}}0.16\\ (0.10)\end{tabular}         & \begin{tabular}[c]{@{}l@{}}-0.03\\ (0.03)\end{tabular}         & \begin{tabular}[c]{@{}l@{}}$0.40^{***}$\\ (0.08)\end{tabular}  & \begin{tabular}[c]{@{}l@{}}-0.06\\ (0.08)\end{tabular}         & \begin{tabular}[c]{@{}l@{}}-0.02\\ (0.05)\end{tabular}         & \begin{tabular}[c]{@{}l@{}}-0.04\\ (0.09)\end{tabular}       & \begin{tabular}[c]{@{}l@{}}0.04\\ (0.09)\end{tabular}               & \begin{tabular}[c]{@{}l@{}}0.01\\ (0.03)\end{tabular}          & \begin{tabular}[c]{@{}l@{}}0.02\\ (0.04)\end{tabular}          \\
{\tt Statuses    }         & \begin{tabular}[c]{@{}l@{}}0.02\\ (0.05)\end{tabular}         & \begin{tabular}[c]{@{}l@{}}-0.03\\ (0.05)\end{tabular}        & \begin{tabular}[c]{@{}l@{}}-0.13\\ (0.11)\end{tabular}       & \begin{tabular}[c]{@{}l@{}}-0.00\\ (0.03)\end{tabular}       & \begin{tabular}[c]{@{}l@{}}$0.21^{**}$\\ (0.07)\end{tabular}   & \begin{tabular}[c]{@{}l@{}}0.04\\ (0.03)\end{tabular}          & \begin{tabular}[c]{@{}l@{}}-0.04\\ (0.11)\end{tabular}        & \begin{tabular}[c]{@{}l@{}}-$0.25^{**}$\\ (0.10)\end{tabular} & \begin{tabular}[c]{@{}l@{}}$0.16^{***}$\\ (0.03)\end{tabular}  & \begin{tabular}[c]{@{}l@{}}-$0.20^{*}$\\ (0.08)\end{tabular}   & \begin{tabular}[c]{@{}l@{}}$0.21^{*}$\\ (0.09)\end{tabular}    & \begin{tabular}[c]{@{}l@{}}$0.18^{**}$\\ (0.05)\end{tabular}   & \begin{tabular}[c]{@{}l@{}}-$0.21^{*}$\\ (0.10)\end{tabular} & \begin{tabular}[c]{@{}l@{}}-0.11\\ (0.09)\end{tabular}              & \begin{tabular}[c]{@{}l@{}}-$0.21^{***}$\\ (0.03)\end{tabular} & \begin{tabular}[c]{@{}l@{}}-$0.12^{**}$\\ (0.04)\end{tabular}  \\
Income               & \begin{tabular}[c]{@{}l@{}}9.85e-6\\ (5.05e-6)\end{tabular}   & \begin{tabular}[c]{@{}l@{}}3.16e-6\\ (5.73e-6)\end{tabular}   & \begin{tabular}[c]{@{}l@{}}-2.11e-5\\ (1.51e-5)\end{tabular} & \begin{tabular}[c]{@{}l@{}}-2.05e-6\\ (3.86e-6)\end{tabular} & \begin{tabular}[c]{@{}l@{}}1.37e-5\\ (7.36e-6)\end{tabular}    & \begin{tabular}[c]{@{}l@{}}5.05e-6\\ (3.99e-6)\end{tabular}    & \begin{tabular}[c]{@{}l@{}}1.49e-5\\ (1.18e-5)\end{tabular}   & \begin{tabular}[c]{@{}l@{}}7.99e-6\\ (1.21e-5)\end{tabular}   & \begin{tabular}[c]{@{}l@{}}-7.43e-6\\ (4.00e-6)\end{tabular}   & \begin{tabular}[c]{@{}l@{}}-4.46e-6\\ (1.03e-5)\end{tabular}   & \begin{tabular}[c]{@{}l@{}}5.75e-6\\ (1.06e-5)\end{tabular}    & \begin{tabular}[c]{@{}l@{}}-1.37e-5\\ (7.66e-6)\end{tabular}   & \begin{tabular}[c]{@{}l@{}}9.89e-6\\ (1.17e-5)\end{tabular}  & \begin{tabular}[c]{@{}l@{}}-$2.98e-5.^{*}$\\ (1.51e-5)\end{tabular} & \begin{tabular}[c]{@{}l@{}}2.92e-6\\ (3.92e-6)\end{tabular}    & \begin{tabular}[c]{@{}l@{}}-7.47e-6\\ (5.42e-6)\end{tabular}   \\
Religious            & \begin{tabular}[c]{@{}l@{}}-$1.25^{**}$\\ (0.42)\end{tabular} & \begin{tabular}[c]{@{}l@{}}-$1.37^{**}$\\ (0.51)\end{tabular} & \begin{tabular}[c]{@{}l@{}}-0.23\\ (0.72)\end{tabular}       & \begin{tabular}[c]{@{}l@{}}0.30\\ (0.17)\end{tabular}        & \begin{tabular}[c]{@{}l@{}}$0.81^{**}$\\ (0.27)\end{tabular}   & \begin{tabular}[c]{@{}l@{}}-0.28\\ (0.21)\end{tabular}         & \begin{tabular}[c]{@{}l@{}}0.48\\ (0.53)\end{tabular}         & \begin{tabular}[c]{@{}l@{}}-0.05\\ (0.73)\end{tabular}        & \begin{tabular}[c]{@{}l@{}}0.27\\ (0.17)\end{tabular}          & \begin{tabular}[c]{@{}l@{}}-0.31\\ (0.59)\end{tabular}         & \begin{tabular}[c]{@{}l@{}}$0.78^{*}$\\ (0.36)\end{tabular}    & \begin{tabular}[c]{@{}l@{}}-0.64\\ (0.47)\end{tabular}         & \begin{tabular}[c]{@{}l@{}}0.22\\ (0.60)\end{tabular}        & \begin{tabular}[c]{@{}l@{}}-0.47\\ (0.72)\end{tabular}              & \begin{tabular}[c]{@{}l@{}}-0.45\\ (0.23)\end{tabular}         & \begin{tabular}[c]{@{}l@{}}$0.76^{***}$\\ (0.20)\end{tabular}  \\
Family               & \begin{tabular}[c]{@{}l@{}}$0.32^{*}$\\ (0.15)\end{tabular}   & \begin{tabular}[c]{@{}l@{}}$0.39^{*}$\\ (0.16)\end{tabular}   & \begin{tabular}[c]{@{}l@{}}0.56\\ (0.37)\end{tabular}        & \begin{tabular}[c]{@{}l@{}}0.13\\ (0.12)\end{tabular}        & \begin{tabular}[c]{@{}l@{}}-0.47\\ (0.26)\end{tabular}         & \begin{tabular}[c]{@{}l@{}}0.03\\ (0.13)\end{tabular}          & \begin{tabular}[c]{@{}l@{}}0.16\\ (0.39)\end{tabular}         & \begin{tabular}[c]{@{}l@{}}-1.02\\ (0.72)\end{tabular}        & \begin{tabular}[c]{@{}l@{}}-0.27\\ (0.14)\end{tabular}         & \begin{tabular}[c]{@{}l@{}}-0.38\\ (0.43)\end{tabular}         & \begin{tabular}[c]{@{}l@{}}0.54\\ (0.28)\end{tabular}          & \begin{tabular}[c]{@{}l@{}}-$0.69^{*}$\\ (0.35)\end{tabular}   & \begin{tabular}[c]{@{}l@{}}0.41\\ (0.37)\end{tabular}        & \begin{tabular}[c]{@{}l@{}}$0.69^{*}$\\ (0.34)\end{tabular}         & \begin{tabular}[c]{@{}l@{}}-0.09\\ (0.14)\end{tabular}         & \begin{tabular}[c]{@{}l@{}}-0.23\\ (0.18)\end{tabular}         \\
Following Trump      & \begin{tabular}[c]{@{}l@{}}-0.03\\ (0.18)\end{tabular}        & \begin{tabular}[c]{@{}l@{}}-0.12\\ (0.21)\end{tabular}        & \begin{tabular}[c]{@{}l@{}}-1.15\\ (0.72)\end{tabular}       & \begin{tabular}[c]{@{}l@{}}-0.27\\ (0.15)\end{tabular}       & \begin{tabular}[c]{@{}l@{}}$1.33^{***}$\\ (0.19)\end{tabular}  & \begin{tabular}[c]{@{}l@{}}-0.28\\ (0.16)\end{tabular}         & \begin{tabular}[c]{@{}l@{}}$0.76^{*}$\\ (0.33)\end{tabular}   & \begin{tabular}[c]{@{}l@{}}0.26\\ (0.47)\end{tabular}         & \begin{tabular}[c]{@{}l@{}}0.04\\ (0.13)\end{tabular}          & \begin{tabular}[c]{@{}l@{}}0.35\\ (0.33)\end{tabular}          & \begin{tabular}[c]{@{}l@{}}$1.13^{***}$\\ (0.24)\end{tabular}  & \begin{tabular}[c]{@{}l@{}}-0.18\\ (0.30)\end{tabular}         & \begin{tabular}[c]{@{}l@{}}-1.53\\ (1.01)\end{tabular}       & \begin{tabular}[c]{@{}l@{}}$0.76^{*}$\\ (0.32)\end{tabular}         & \begin{tabular}[c]{@{}l@{}}-0.11\\ (0.15)\end{tabular}         & \begin{tabular}[c]{@{}l@{}}-$0.80^{**}$\\ (0.24)\end{tabular}  \\
Following Biden      & \begin{tabular}[c]{@{}l@{}}$0.73^{***}$\\ (0.12)\end{tabular} & \begin{tabular}[c]{@{}l@{}}$0.54^{***}$\\ (0.13)\end{tabular} & \begin{tabular}[c]{@{}l@{}}-0.45\\ (0.39)\end{tabular}       & \begin{tabular}[c]{@{}l@{}}$0.29^{**}$\\ (0.09)\end{tabular} & \begin{tabular}[c]{@{}l@{}}-$2.03^{***}$\\ (0.37)\end{tabular} & \begin{tabular}[c]{@{}l@{}}$0.39^{***}$\\ (0.09)\end{tabular}  & \begin{tabular}[c]{@{}l@{}}-$1.51^{*}$\\ (0.60)\end{tabular}  & \begin{tabular}[c]{@{}l@{}}-0.88\\ (0.47)\end{tabular}        & \begin{tabular}[c]{@{}l@{}}-$0.26^{*}$\\ (0.10)\end{tabular}   & \begin{tabular}[c]{@{}l@{}}-$0.66^{*}$\\ (0.31)\end{tabular}   & \begin{tabular}[c]{@{}l@{}}-$2.53^{***}$\\ (0.72)\end{tabular} & \begin{tabular}[c]{@{}l@{}}-$0.57^{*}$\\ (0.24)\end{tabular}   & \begin{tabular}[c]{@{}l@{}}-0.55\\ (0.37)\end{tabular}       & \begin{tabular}[c]{@{}l@{}}-0.42\\ (0.35)\end{tabular}              & \begin{tabular}[c]{@{}l@{}}-$0.49^{***}$\\ (0.11)\end{tabular} & \begin{tabular}[c]{@{}l@{}}$0.37^{**}$\\ (0.12)\end{tabular}   \\
Urban                & \begin{tabular}[c]{@{}l@{}}-0.13\\ (0.13)\end{tabular}        & \begin{tabular}[c]{@{}l@{}}-0.05\\ (0.15)\end{tabular}        & \begin{tabular}[c]{@{}l@{}}-0.07\\ (0.32)\end{tabular}       & \begin{tabular}[c]{@{}l@{}}0.03\\ (0.10)\end{tabular}        & \begin{tabular}[c]{@{}l@{}}-$0.97^{***}$\\ (0.17)\end{tabular} & \begin{tabular}[c]{@{}l@{}}0.20\\ (0.11)\end{tabular}          & \begin{tabular}[c]{@{}l@{}}-0.24\\ (0.32)\end{tabular}        & \begin{tabular}[c]{@{}l@{}}0.35\\ (0.38)\end{tabular}         & \begin{tabular}[c]{@{}l@{}}0.18\\ (0.10)\end{tabular}          & \begin{tabular}[c]{@{}l@{}}$0.77^{*}$\\ (0.32)\end{tabular}    & \begin{tabular}[c]{@{}l@{}}-0.33\\ (0.23)\end{tabular}         & \begin{tabular}[c]{@{}l@{}}0.07\\ (0.18)\end{tabular}          & \begin{tabular}[c]{@{}l@{}}-0.03\\ (0.32)\end{tabular}       & \begin{tabular}[c]{@{}l@{}}-0.07\\ (0.29)\end{tabular}              & \begin{tabular}[c]{@{}l@{}}0.09\\ (0.11)\end{tabular}          & \begin{tabular}[c]{@{}l@{}}-0.10\\ (0.12)\end{tabular}         \\
Suburban             & \begin{tabular}[c]{@{}l@{}}-0.16\\ (0.17)\end{tabular}        & \begin{tabular}[c]{@{}l@{}}0.12\\ (0.19)\end{tabular}         & \begin{tabular}[c]{@{}l@{}}0.46\\ (0.39)\end{tabular}        & \begin{tabular}[c]{@{}l@{}}-0.00\\ (0.13)\end{tabular}       & \begin{tabular}[c]{@{}l@{}}-0.41\\ (0.22)\end{tabular}         & \begin{tabular}[c]{@{}l@{}}0.21\\ (0.14)\end{tabular}          & \begin{tabular}[c]{@{}l@{}}0.52\\ (0.37)\end{tabular}         & \begin{tabular}[c]{@{}l@{}}0.19\\ (0.51)\end{tabular}         & \begin{tabular}[c]{@{}l@{}}0.01\\ (0.13)\end{tabular}          & \begin{tabular}[c]{@{}l@{}}0.54\\ (0.41)\end{tabular}          & \begin{tabular}[c]{@{}l@{}}-0.45\\ (0.33)\end{tabular}         & \begin{tabular}[c]{@{}l@{}}0.31\\ (0.23)\end{tabular}          & \begin{tabular}[c]{@{}l@{}}-0.90\\ (0.58)\end{tabular}       & \begin{tabular}[c]{@{}l@{}}-0.69\\ (0.48)\end{tabular}              & \begin{tabular}[c]{@{}l@{}}0.11\\ (0.14)\end{tabular}          & \begin{tabular}[c]{@{}l@{}}-0.15\\ (0.17)\end{tabular}         \\
{\tt White }               & \begin{tabular}[c]{@{}l@{}}$0.49^{**}$\\ (0.15)\end{tabular}  & \begin{tabular}[c]{@{}l@{}}$0.66^{***}$\\ (0.17)\end{tabular} & \begin{tabular}[c]{@{}l@{}}-0.22\\ (0.27)\end{tabular}       & \begin{tabular}[c]{@{}l@{}}-0.08\\ (0.09)\end{tabular}       & \begin{tabular}[c]{@{}l@{}}$0.51^{*}$\\ (0.22)\end{tabular}    & \begin{tabular}[c]{@{}l@{}}-0.01\\ (0.10)\end{tabular}         & \begin{tabular}[c]{@{}l@{}}$0.92^{*}$\\ (0.39)\end{tabular}   & \begin{tabular}[c]{@{}l@{}}-0.39\\ (0.24)\end{tabular}        & \begin{tabular}[c]{@{}l@{}}-0.05\\ (0.09)\end{tabular}         & \begin{tabular}[c]{@{}l@{}}-0.00\\ (0.20)\end{tabular}         & \begin{tabular}[c]{@{}l@{}}$0.69^{*}$\\ (0.31)\end{tabular}    & \begin{tabular}[c]{@{}l@{}}-$0.34^{*}$\\ (0.14)\end{tabular}   & \begin{tabular}[c]{@{}l@{}}0.44\\ (0.29)\end{tabular}        & \begin{tabular}[c]{@{}l@{}}-0.49\\ (0.25)\end{tabular}              & \begin{tabular}[c]{@{}l@{}}-$0.29^{**}$\\ (0.08)\end{tabular}  & \begin{tabular}[c]{@{}l@{}}0.22\\ (0.12)\end{tabular}          \\
{\tt Black }               & \begin{tabular}[c]{@{}l@{}}0.29\\ (0.19)\end{tabular}         & \begin{tabular}[c]{@{}l@{}}0.35\\ (0.22)\end{tabular}         & \begin{tabular}[c]{@{}l@{}}-0.62\\ (0.41)\end{tabular}       & \begin{tabular}[c]{@{}l@{}}-0.10\\ (0.11)\end{tabular}       & \begin{tabular}[c]{@{}l@{}}-0.56\\ (0.35)\end{tabular}         & \begin{tabular}[c]{@{}l@{}}-0.06\\ (0.13)\end{tabular}         & \begin{tabular}[c]{@{}l@{}}-0.55\\ (0.69)\end{tabular}        & \begin{tabular}[c]{@{}l@{}}-$1.12^{*}$\\ (0.45)\end{tabular}  & \begin{tabular}[c]{@{}l@{}}$0.63^{***}$\\ (0.10)\end{tabular}  & \begin{tabular}[c]{@{}l@{}}-0.13\\ (0.27)\end{tabular}         & \begin{tabular}[c]{@{}l@{}}-0.51\\ (0.48)\end{tabular}         & \begin{tabular}[c]{@{}l@{}}0.14\\ (0.16)\end{tabular}          & \begin{tabular}[c]{@{}l@{}}0.02\\ (0.42)\end{tabular}        & \begin{tabular}[c]{@{}l@{}}-0.38\\ (0.34)\end{tabular}              & \begin{tabular}[c]{@{}l@{}}-$0.43^{***}$\\ (0.12)\end{tabular} & \begin{tabular}[c]{@{}l@{}}-0.15\\ (0.16)\end{tabular}         \\
{\tt Hispanic  }           & \begin{tabular}[c]{@{}l@{}}0.62\\ (0.40)\end{tabular}         & \begin{tabular}[c]{@{}l@{}}0.80\\ (0.43)\end{tabular}         & \begin{tabular}[c]{@{}l@{}}0.13\\ (0.75)\end{tabular}        & \begin{tabular}[c]{@{}l@{}}-0.38\\ (0.32)\end{tabular}       & \begin{tabular}[c]{@{}l@{}}0.35\\ (0.59)\end{tabular}          & \begin{tabular}[c]{@{}l@{}}-0.61\\ (0.38)\end{tabular}         & -                                                             & -                                                             & \begin{tabular}[c]{@{}l@{}}0.39\\ (0.24)\end{tabular}          & \begin{tabular}[c]{@{}l@{}}0.31\\ (0.54)\end{tabular}          & \begin{tabular}[c]{@{}l@{}}0.43\\ (0.79)\end{tabular}          & \begin{tabular}[c]{@{}l@{}}0.24\\ (0.35)\end{tabular}          & \begin{tabular}[c]{@{}l@{}}0.10\\ (1.05)\end{tabular}        & \begin{tabular}[c]{@{}l@{}}0.47\\ (0.63)\end{tabular}               & \begin{tabular}[c]{@{}l@{}}-0.56\\ (0.32)\end{tabular}         & \begin{tabular}[c]{@{}l@{}}-0.07\\ (0.41)\end{tabular}         \\
Hate crimes          & \begin{tabular}[c]{@{}l@{}}-0.55\\ (0.51)\end{tabular}        & \begin{tabular}[c]{@{}l@{}}0.32\\ (0.53)\end{tabular}         & \begin{tabular}[c]{@{}l@{}}-1.08\\ (1.30)\end{tabular}       & \begin{tabular}[c]{@{}l@{}}-0.36\\ (0.35)\end{tabular}       & \begin{tabular}[c]{@{}l@{}}-$1.95^{*}$\\ (0.84)\end{tabular}   & \begin{tabular}[c]{@{}l@{}}0.22\\ (0.37)\end{tabular}          & \begin{tabular}[c]{@{}l@{}}0.27\\ (1.21)\end{tabular}         & \begin{tabular}[c]{@{}l@{}}1.24\\ (1.04)\end{tabular}         & \begin{tabular}[c]{@{}l@{}}0.28\\ (0.35)\end{tabular}          & \begin{tabular}[c]{@{}l@{}}-0.32\\ (0.89)\end{tabular}         & \begin{tabular}[c]{@{}l@{}}-0.14\\ (1.01)\end{tabular}         & \begin{tabular}[c]{@{}l@{}}-0.80\\ (0.67)\end{tabular}         & \begin{tabular}[c]{@{}l@{}}$3.15^{**}$\\ (0.94)\end{tabular} & \begin{tabular}[c]{@{}l@{}}0.29\\ (1.12)\end{tabular}               & \begin{tabular}[c]{@{}l@{}}0.15\\ (0.36)\end{tabular}          & \begin{tabular}[c]{@{}l@{}}0.28\\ (0.45)\end{tabular}          \\
N                    & \multicolumn{16}{c}{6,768}          \\
\bottomrule
\end{tabular}
\begin{tablenotes}
    \small
    \item Note. * $p<0.05$. ** $p<0.01$. *** $p<0.001$. Table entries are coefficients (standard errors).
    \end{tablenotes}
    \end{threeparttable}
\end{sidewaystable*}

\begin{table*}[htbp!]

\centering
\begin{threeparttable}
\caption{Logistic regression outputs for the level 2 topics of ``Double standard'' on the \#StopAsianHate and \#StopAAPIHate movement against demographics and other variables of interest.}
    \label{tab:level2_logit_double}
    \centering
\begin{tabular}{llll}
\toprule
Independent variable & DB.1                                                           & DB.2                                                          & DB.3                                                          \\
\midrule
Male                 & \begin{tabular}[c]{@{}l@{}}-0.20\\ (0.17)\end{tabular}         & \begin{tabular}[c]{@{}l@{}}0.38\\ (0.20)\end{tabular}         & \begin{tabular}[c]{@{}l@{}}-0.33\\ (0.40)\end{tabular}        \\
Age (years)                & \begin{tabular}[c]{@{}l@{}}-$0.02^{***}$\\ (0.001\end{tabular} & \begin{tabular}[c]{@{}l@{}}0.01\\ (0.01)\end{tabular}         & \begin{tabular}[c]{@{}l@{}}$0.05^{***}$\\ (0.01)\end{tabular} \\
{\tt Verified  }           & -                                                              & -                                                             & -                                                             \\
{\tt Followers   }         & \begin{tabular}[c]{@{}l@{}}-0.21\\ (0.12)\end{tabular}         & \begin{tabular}[c]{@{}l@{}}0.12\\ (0.13)\end{tabular}         & \begin{tabular}[c]{@{}l@{}}0.23\\ (0.27)\end{tabular}         \\
{\tt Friends     }         & \begin{tabular}[c]{@{}l@{}}0.01\\ (0.11)\end{tabular}          & \begin{tabular}[c]{@{}l@{}}-0.15\\ (0.12)\end{tabular}        & \begin{tabular}[c]{@{}l@{}}0.49\\ (0.27)\end{tabular}         \\
{\tt List memberships }    & \begin{tabular}[c]{@{}l@{}}$0.47^{**}$\\ (0.18)\end{tabular}   & \begin{tabular}[c]{@{}l@{}}-0.28\\ (0.20)\end{tabular}        & \begin{tabular}[c]{@{}l@{}}-$0.77^{*}$\\ (0.36)\end{tabular}  \\
{\tt Favourites     }      & \begin{tabular}[c]{@{}l@{}}$0.12^{*}$\\ (0.06)\end{tabular}    & \begin{tabular}[c]{@{}l@{}}-$0.13^{*}$\\ (0.06)\end{tabular}  & \begin{tabular}[c]{@{}l@{}}-0.06\\ (0.17)\end{tabular}        \\
{\tt Statuses   }          & \begin{tabular}[c]{@{}l@{}}-0.09\\ (0.08)\end{tabular}         & \begin{tabular}[c]{@{}l@{}}0.06\\ (0.09)\end{tabular}         & \begin{tabular}[c]{@{}l@{}}0.17\\ (0.22)\end{tabular}         \\
Income               & \begin{tabular}[c]{@{}l@{}}-1.73e-5\\ (9.19e-6)\end{tabular}   & \begin{tabular}[c]{@{}l@{}}1.70e-5\\ (1.01e-5)\end{tabular}   & \begin{tabular}[c]{@{}l@{}}2.43e-5\\ (2.32e-5)\end{tabular}   \\
Religious            & \begin{tabular}[c]{@{}l@{}}-0.25\\ (0.39)\end{tabular}         & \begin{tabular}[c]{@{}l@{}}-0.71\\ (0.73)\end{tabular}        & \begin{tabular}[c]{@{}l@{}}1.00\\ (0.52)\end{tabular}         \\
Family               & \begin{tabular}[c]{@{}l@{}}-0.51\\ (0.34)\end{tabular}         & \begin{tabular}[c]{@{}l@{}}0.11\\ (0.48)\end{tabular}         & \begin{tabular}[c]{@{}l@{}}0.93\\ (0.51)\end{tabular}         \\
Following Trump      & \begin{tabular}[c]{@{}l@{}}-0.08\\ (0.34)\end{tabular}         & \begin{tabular}[c]{@{}l@{}}-0.73\\ (0.60)\end{tabular}        & \begin{tabular}[c]{@{}l@{}}0.44\\ (0.48)\end{tabular}         \\
Following Biden      & \begin{tabular}[c]{@{}l@{}}-0.14\\ (0.37)\end{tabular}         & \begin{tabular}[c]{@{}l@{}}0.49\\ (0.40)\end{tabular}         & \begin{tabular}[c]{@{}l@{}}-1.20\\ (1.09)\end{tabular}        \\
Urban                & \begin{tabular}[c]{@{}l@{}}-0.08\\ (0.25)\end{tabular}         & \begin{tabular}[c]{@{}l@{}}0.17\\ (0.30)\end{tabular}         & \begin{tabular}[c]{@{}l@{}}-0.09\\ (0.48)\end{tabular}        \\
Suburban             & \begin{tabular}[c]{@{}l@{}}0.15\\ (0.34)\end{tabular}          & \begin{tabular}[c]{@{}l@{}}-0.01\\ (0.39)\end{tabular}        & \begin{tabular}[c]{@{}l@{}}-0.39\\ (0.73)\end{tabular}        \\
{\tt White    }            & \begin{tabular}[c]{@{}l@{}}0.15\\ (0.28)\end{tabular}          & \begin{tabular}[c]{@{}l@{}}-0.47\\ (0.34)\end{tabular}        & \begin{tabular}[c]{@{}l@{}}0.28\\ (0.52)\end{tabular}         \\
{\tt Black      }          & \begin{tabular}[c]{@{}l@{}}-$0.98^{***}$\\ (0.26)\end{tabular} & \begin{tabular}[c]{@{}l@{}}$1.24^{***}$\\ (0.29)\end{tabular} & \begin{tabular}[c]{@{}l@{}}-1.41\\ (0.86)\end{tabular}        \\
{\tt Hispanic       }      & -                                                              & -                                                             & -                                                             \\
Hate crimes          & \begin{tabular}[c]{@{}l@{}}0.56\\ (0.97)\end{tabular}          & \begin{tabular}[c]{@{}l@{}}0.18\\ (1.05)\end{tabular}         & \begin{tabular}[c]{@{}l@{}}-4.15\\ (2.49)\end{tabular}        \\
N                    & \multicolumn{3}{c}{3,856}    \\
\bottomrule
\end{tabular}
\begin{tablenotes}
    \small
    \item Note. * $p<0.05$. ** $p<0.01$. *** $p<0.001$. Table entries are coefficients (standard errors).
    \end{tablenotes}
    \end{threeparttable}
\end{table*}

\begin{table*}[htbp!]

\centering
\begin{threeparttable}
\caption{Logistic regression outputs for the level 2 topics of ``Policy'' on the \#StopAsianHate and \#StopAAPIHate movement against demographics and other variables of interest.}
    \label{tab:level2_logit_policy}
    \centering
\begin{tabular}{ll}
\toprule
Independent variable & P.1                                                          \\
\midrule
Male                 & \begin{tabular}[c]{@{}l@{}}0.23\\ (0.25)\end{tabular}        \\
Age (years)                 & \begin{tabular}[c]{@{}l@{}}-0.01\\ (0.01)\end{tabular}       \\
{\tt Verified   }          & -                                                            \\
{\tt Followers }          & \begin{tabular}[c]{@{}l@{}}0.28\\ (0.16)\end{tabular}        \\
{\tt Friends    }          & \begin{tabular}[c]{@{}l@{}}-$0.31^{*}$\\ (0.15)\end{tabular} \\
{\tt List memberships}     & \begin{tabular}[c]{@{}l@{}}0.09\\ (0.23)\end{tabular}        \\
{\tt Favourites}           & \begin{tabular}[c]{@{}l@{}}$0.22^{*}$\\ (0.10)\end{tabular}  \\
{\tt Statuses   }          & \begin{tabular}[c]{@{}l@{}}-0.05\\ (0.12)\end{tabular}       \\
Income               & \begin{tabular}[c]{@{}l@{}}1.98e-5\\ (1.39e-5)\end{tabular}  \\
Religious            & \begin{tabular}[c]{@{}l@{}}-0.63\\ (0.73)\end{tabular}       \\
Family               & \begin{tabular}[c]{@{}l@{}}-0.45\\ (0.36)\end{tabular}       \\
Following Trump      & \begin{tabular}[c]{@{}l@{}}1.07\\ (0.76)\end{tabular}        \\
Following Biden      & \begin{tabular}[c]{@{}l@{}}0.00\\ (0.28)\end{tabular}        \\
Urban                & \begin{tabular}[c]{@{}l@{}}0.52\\ (0.32)\end{tabular}        \\
Suburban             & \begin{tabular}[c]{@{}l@{}}-0.40\\ (0.39)\end{tabular}       \\
{\tt White   }             & \begin{tabular}[c]{@{}l@{}}0.10\\ (0.37)\end{tabular}        \\
{\tt Black  }              & \begin{tabular}[c]{@{}l@{}}-$0.86^{*}$\\ (0.43)\end{tabular} \\
{\tt Hispanic}            & \begin{tabular}[c]{@{}l@{}}0.06\\ (1.12)\end{tabular}        \\
Hate crimes          & \begin{tabular}[c]{@{}l@{}}2.85\\ (1.48)\end{tabular}        \\
N                    & 719          \\
\bottomrule
\end{tabular}

    \begin{tablenotes}
    \small
    \item Note. * $p<0.05$. ** $p<0.01$. *** $p<0.001$. Table entries are coefficients (standard errors).
    \end{tablenotes}
    \end{threeparttable}
\end{table*}

\end{document}